\documentclass[sigconf]{acmart}
\usepackage{tikz}
\usetikzlibrary{patterns}
\usepackage{subcaption}
\usepackage{algorithm}
\usepackage{algpseudocode}
\usepackage{amsmath}
\usepackage{graphicx} 
\usepackage{caption}
\usepackage{diagbox}
\usepackage{hyperref}
\DeclareMathOperator*{\argmin}{\arg\!\min}

\algtext*{EndFor}
\algtext*{EndIf}
\algtext*{EndFunction}
\usepackage{booktabs} 
\newcommand{\eps}{\varepsilon}
\newcommand{\epss}{\frac{1}{\varepsilon}}
\newcommand{\etal}{\textit{et al}.}
\newtheorem{theorem}{Theorem}[section]

\newtheorem{lemma}[theorem]{Lemma}

\setcopyright{none}
\def\BibTeX{{\rm B\kern-.05em{\sc i\kern-.025em b}\kern-.08emT\kern-.1667em\lower.7ex\hbox{E}\kern-.125emX}}
    
\begin{document}

%

\title{Streaming Quantiles Algorithms with~Small~Space~and~Update~Time}

%
\author{Nikita Ivkin}
\email{ivkin@amazon.com}
\affiliation{%
  \institution{Amazon}
}
\author{Edo Liberty}
\email{libertye@amazon.com}
\affiliation{%
  \institution{Amazon}
}
\author{Kevin Lang}
\email{langk@verizonmedia.com}
\affiliation{\institution{Yahoo Research}
}

\author{Zohar Karnin}
\email{zkarnin@amazon.com}
\affiliation{%
  \institution{Amazon}
}
\author{Vladimir Braverman}
\email{vova@cs.jhu.edu}
\affiliation{
\institution{Johns Hopkins University}
}

%
%
%
%
%
%
%
%
\renewcommand{\shortauthors}{Ivkin, Liberty, Lang, Karnin and Braverman}

\begin{abstract}
	Approximating quantiles and distributions over streaming data has been studied for roughly two decades now.
	Recently, Karnin, Lang, and Liberty 
	proposed the first asymptotically optimal algorithm for doing so. 
	This manuscript complements their theoretical result by providing a practical variants of their algorithm with improved constants. 
	For a given sketch size, our techniques provably reduce the upper bound on the sketch error by a factor of two.
	These improvements are verified experimentally.
	Our modified quantile sketch improves the latency as well by reducing the worst-case update time from $O(\epss)$ down to $O(\log{\epss})$. 
	We also suggest two algorithms for weighted item streams which offer improved asymptotic update times compared to na\"ive extensions.  
	Finally, we provide a specialized data structure for these sketches which reduces both their memory footprints and update times.
\end{abstract}

%
%

%

%
%
\maketitle

\section{Introduction}
Estimating the underlying distribution of data is crucial for many applications.
It is common to approximate an entire Cumulative Distribution Function (CDF) or specific quantiles. 
The median ($0.5$ quantile) and $95$-th and $99$-th percentiles are widely used in financial metrics, statistical tests, and system monitoring.
Quantiles summary found applications in databases~\cite{selinger1979access, poosala1996improved}, sensor networks~\cite{li2011ubiquitous}, 
logging systems~\cite{pike2005interpreting}, distributed systems~\cite{dewitt1991parallel}, and decision trees~\cite{chen2016xgboost}.
While computing quantiles is conceptually very simple, doing so naively becomes infeasible for very large data.

Formally the quantiles problem can be defined as follows. 
Let $S$ be a multiset of items $S = \left\{ s_i \right\}_{i=1}^n$.
The items in $S$ exhibit a full-ordering and the corresponding smaller-than comparator is known.
The rank of a query $q$ (w.r.t.\ $S$) is the number of items in $S$ which are smaller than $q$.
An algorithm should process $S$ such that it can compute the rank of any query item.
Answering rank queries exactly for every query is trivially possible by storing the multiset $S$. 
Storing $S$ in its entirety is also necessary for this task.  

An approximate version of the problem relaxes this requirement.
It is allowed to output an approximate rank which is off by at most $\eps n$ from the exact rank.
In a randomized setting, the algorithm is allowed to fail with probability at most $\delta$. 
Note that, for the randomized version to provide a correct answer to 
all possible queries, it suffices to amplify the success probability by running the algorithm with a failure probability of $\delta\eps$, and  
applying the union bound over $O(\epss)$ quantiles.
Uniform random sampling of $O(\frac{1}{\eps^2}\log\frac{1}{\delta\eps})$ solves this problem.

In network monitoring \cite{liu2016one} and other applications it is critical to 
maintain statistics while making only a single pass over the data and minimizing the communication and update time. 
As a result, the problem of approximating quantiles was considered in several models including
distributed settings~\cite{dewitt1991parallel, greenwald2004power, shrivastava2004medians}, 
continuous monitoring~\cite{cormode2005holistic,yi2013optimal}, 
streaming~\cite{agarwal2013mergeable, karnin2016optimal, manku1998approximate, manku1999random, wang2013quantiles, greenwald2001space, greenwald2016quantiles}, 
and sliding windows~\cite{arasu2004approximate, lin2004continuously}. 
In the present paper, the quantiles problem is considered in a standard streaming setting.
The algorithm receives the items in $S$ one by one in an iterative manner.
The algorithm's approximation guarantees should not depend on the order or 
the content of the updates $s_t$, and its space complexity should depend on $n$ at most poly-logarithmically.\footnote{Throughout this manuscript we assume that each item in the stream requires $O(1)$ space to store.}

In their pioneering paper~\cite{munro1980selection}, Munro and Paterson showed that one would need $\Omega(n^{1/p})$ space and $p$ passes over the dataset to find a median. 
They also suggested an optimal iterative algorithm to find it. Later Manku~\etal~\cite{manku1998approximate} showed that the first iteration 
of the algorithm in~\cite{munro1980selection} can be used to solve the $\eps$ approximate quantile problem in one pass using only 
$O(\epss \log^2 n)$ space. Note that, for a small enough $\eps$, this is a significant improvement over the naive algorithm, 
which samples $O(\frac{1}{\eps^2}\log{\epss})$ items of the stream using reservoir sampling. 
The algorithm in ~\cite{manku1998approximate} is deterministic, however, compared with reservoir sampling it assumes the length of the 
stream is known in advance. In many applications such an assumption is unrealistic. In their follow-up paper~\cite{manku1999random} the authors suggested 
a randomized algorithm without that assumption. 
Further improvement by Agarwal~\etal~\cite{agarwal2013mergeable} via randomizing the core subroutine pushed the space requirements down to $O(\epss\log^{3/2}{\epss})$. 
Also, the new data structure was proven to be fully mergeable.  
Greenwald and Khanna in~\cite{greenwald2001space} presented an algorithm  that maintains upper and lower bounds for each quantile individually, rather
than one bound for all quantiles. It is deterministic and requires only $O(\epss \log\eps n)$ space. It is not known to be fully mergeable. 
Later Felber and Ostrovsky~\cite{felber2015randomized} suggested non-trivial techniques of feeding sampled items into sketches 
from~\cite{greenwald2001space} and improved the space complexity to $O(\epss \log \epss)$.
Recently Karnin~\etal~in~\cite{karnin2016optimal} presented an asymptotically optimal but non-mergeable data structure with space usage of 
$O(\epss \log\log \epss)$ and a matching lower bound.
They also presented a fully mergeable algorithm whose space complexity is $O(\epss \log^2\log \epss)$.

In the current paper, we suggest several further improvements to the algorithms introduced in~\cite{karnin2016optimal}. 
These improvements do not affect the asymptotic guarantees of~\cite{karnin2016optimal} but reduce the upper bounds by constant terms, both in theory and practice. 
The suggested techniques also improve the worst-case update time. 
Additionally, we suggest two algorithms for the extended version of the problem where updates have weights. 
All the algorithms presented operate in the comparison model. They can only store (and discard) items from the stream and compare between them.
For more background on quantile algorithms in the streaming model see \cite{greenwald2016quantiles, wang2013quantiles}.    

\section{A unified view of previous randomized solutions}\label{section:unifiedview}
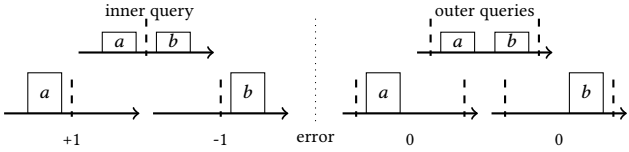
\begin{figure}[!t] 
	\begin{tikzpicture}[scale=0.9]
\def\A{5}
\def\B{-0.9}
\def\C{1.1}

\node[above] at (0.1,-1.05) {\footnotesize $a$};
\draw (-0.15,-1) rectangle (.35,-.7);
\node[above] at (0.9,-1.05) {\footnotesize $b$};
\draw (0.65,-1) rectangle (1.15,-.7);
\draw[->,thick] (-0.5, -1.) -- (1.5,-1);
\node[above ] at (0.55,-0.65) {\footnotesize inner query};
\draw[dashed,thick] (0.5, -0.5) -- (0.5,-1.05);

\node[above] at (0.1 + \A,-1.05) {\footnotesize $a$};
\draw (-0.15 + \A,-1) rectangle (.35 + \A,-.7);
\node[above] at (0.9 + \A,-1.05) {\footnotesize $b$};
\draw (0.65 + \A,-1) rectangle (1.15 + \A,-.7);
\draw[->,thick] (-0.5 + \A, -1) -- (1.5 + \A,-1);
\node[above ] at (0.5 + \A,-0.65) {\footnotesize outer queries};
\draw[dashed,thick] (-0.3 + \A,-0.5) -- (-0.3 + \A,-1.05);
\draw[dashed,thick] (1.3 + \A, -0.5) -- (1.3 + \A,-1.05);

\node[above] at (0.1- \C,-0.9 + \B) {\footnotesize $a$};
\draw (-0.15- \C,-1+ \B) rectangle (.35- \C,-.4+ \B);
\draw[->,thick] (-0.5- \C, -1.+ \B) -- (1.5- \C,-1+ \B);
\draw[dashed,thick] (0.5- \C, -0.5+ \B) -- (0.5- \C,-1.05+ \B);

\node[above] at (0.9+ \C,-0.9 + \B ) {\footnotesize $b$};
\draw (0.65+ \C,-1+ \B) rectangle (1.15+ \C,-.4+ \B);
\draw[->,thick] (-0.5+ \C, -1.+ \B) -- (1.5+ \C,-1+ \B);
\draw[dashed,thick] (0.5+ \C, -0.5+ \B) -- (0.5+ \C,-1.05+ \B);

\node[above] at (0.1- \C + \A,-0.9 + \B) {\footnotesize $a$};
\draw (-0.15- \C + \A,-1+ \B) rectangle (.35- \C + \A,-.4+ \B);
\draw[->,thick] (-0.5- \C + \A, -1.+ \B) -- (1.5- \C + \A,-1+ \B);
\draw[dashed,thick] (-0.3 - \C+ \A,-0.5+ \B) -- (-0.3 - \C+ \A,-1.05+ \B);
\draw[dashed,thick] (1.3 - \C+ \A, -0.5+ \B) -- (1.3 - \C+ \A,-1.05+ \B);

\node[above] at (0.9+ \C + \A,-0.9 + \B ) {\footnotesize $b$};
\draw (0.65+ \C + \A,-1+ \B) rectangle (1.15+ \C + \A,-.4+ \B);
\draw[->,thick] (-0.5+ \C + \A, -1.+ \B) -- (1.5+ \C + \A,-1+ \B);
\draw[dashed,thick] (-0.3 + \C+ \A,-0.5+ \B) -- (-0.3 + \C+ \A,-1.05+ \B);
\draw[dashed,thick] (1.3 + \C+ \A, -0.5+ \B) -- (1.3 + \C+ \A,-1.05+ \B);

\draw[dotted] (0.5 + \A/2, -0.5) -- (0.5 + \A/2, -0.5 + 1.75*\B); 

\node[below] at (0.5 + \A/2,-0.5 + 1.75*\B) {\footnotesize error};
\node[below] at (0.5 - \C,-0.5 + 1.75*\B) {\footnotesize +1};
\node[below] at (0.5 + \C,-0.5 + 1.75*\B) {\footnotesize -1};

\node[below] at (0.5 + \A - \C,-0.5 + 1.75*\B) {\footnotesize 0};
\node[below] at (0.5 + \A + \C,-0.5 + 1.75*\B) {\footnotesize 0};









\end{tikzpicture}
	\vspace*{-10pt}
	\caption{\small {One pair compression for (a,b) introduces $\pm 1$ rank error to 
	inner queries and no error to outer queries.}}\label{scheme:paircompression}
	\vspace*{-15pt}
\end{figure}

To introduce further improvements to the streaming quantiles algorithms we will first re-explain the previous
work using simplified concepts of one pair compression and a compactor. 
Consider a simple problem in which your data set contains only two items $a$ and $b$, while your data structure can only store one item. 
We focus on the comparison based framework where we can only compare items and cannot compute new items via operations such as averaging. 
In this framework, the only option for the data structure is to pick one of them and store it explicitly.
The stored item $x$ is assigned weight~$2$. Given a rank query $q$ the data structure will report $0$ for $q < x$, 
and $2$ for $q > x$. For $q \notin [a, b] $ the output of the data structure will be correct,
however, for $q \in [a,b]$ the correct rank is $1$ and the data structure will output with $0$ or $2$.
It, therefore, introduces a $+1/-1$ error depending on which item was retained. 
From this point on, $q$ is an inner query with respect to the pair $(a,b)$  if $q \in [a,b]$ and an outer query otherwise. 
This lets us distinguish those queries for which an error is introduced from those that were not influenced by a compression. 
Figure~\ref{scheme:paircompression} depicts the above example of \emph{one pair compression}.

The example gives rise to a high-level method for the original problem with a dataset of size $n$ and memory capacity of $k$ items. 
Namely 1) keep adding items to the data structure until it is full; 2) choose any pair of items with the same weight and compress them. 
Notice that if we choose those pairs without care, in the worst case, we might end up representing the full dataset by its top $k$ elements, 
introducing an error of almost $n$ which is much larger than $\eps n$. 
Intuitively, pairs being compacted (compressed) should have their ranks as close as possible, 
thereby affecting as few queries as possible.

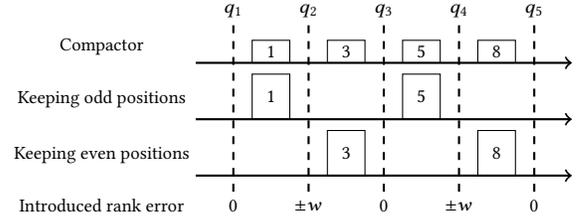
\begin{figure}[!t] 
	\begin{tikzpicture}[scale=1]

\node[align=center,above ] at (-1.75,-1) {\footnotesize Compactor};           
\node[align=center,above ] at (-1.75,0.3-2) {\footnotesize Keeping odd positions};
\node[align=center,above ] at (-1.75,0.55-3) {\footnotesize Keeping even positions};
\node[align=center ] at (-1.75,-2.9) {\footnotesize Introduced rank error};

\node[above ] at (0,-0.5) {\footnotesize $q_1$};
\draw[dashed,thick] (0, -0.5) -- (0,-2.75);
\node[ ] at (0,-2.9) {\footnotesize $0$};

\node[above ] at (1,-0.5) {\footnotesize $q_2$};
\draw[dashed,thick] (1, -0.5) -- (1,-2.75);
\node[ ] at (1,-2.9) {\footnotesize $\pm w$};

\node[above ] at (2,-0.5) {\footnotesize $q_3$};
\draw[dashed,thick] (2, -0.5) -- (2,-2.75);
\node[ ] at (2,-2.9) {\footnotesize $0$};

\node[above ] at (3,-0.5) {\footnotesize $q_4$};
\draw[dashed,thick] (3, -0.5) -- (3,-2.75);
\node[ ] at (3,-2.9) {\footnotesize $\pm w$};

\node[above ] at (4,-0.5) {\footnotesize $q_5$};
\draw[dashed,thick] (4, -0.5) -- (4,-2.75);
\node[ ] at (4,-2.9) {\footnotesize $0$};

\draw[->,thick] (-0.5, -1) -- (4.5,-1);
\draw[->,thick] (-0.5, 0.25-2) -- (4.5,0.25-2);
\draw[->,thick] (-0.5, 0.5-3) -- (4.5,0.5-3);

\draw (0.25,-1) rectangle (.75,-.7) node[pos=.5] {\footnotesize $1$} ; 
\draw (1.25,-1) rectangle (1.75,-.7) node[pos=.5] {\footnotesize $3$} ; 
\draw (2.25,-1) rectangle (2.75,-.7) node[pos=.5] {\footnotesize $5$};
\draw (3.25,-1) rectangle (3.75,-.7) node[pos=.5] {\footnotesize $8$};

\draw (0.25,0.25-2) rectangle (.75,0.25-1.4) node[pos=.5] {\footnotesize $1$} ;
\draw (2.25,0.25-2) rectangle (2.75,0.25-1.4) node[pos=.5] {\footnotesize $5$} ;

\draw (1.25,0.5-3) rectangle (1.75,0.5-2.4) node[pos=.5] {\footnotesize $3$} ;
\draw (3.25,0.5-3) rectangle (3.75,0.5-2.4) node[pos=.5] {\footnotesize $8$} ;

\end{tikzpicture}
	\vspace*{-10pt}
	\caption{\small {Compacting $[1,3,5,8]$ introduces $\pm w$ rank error to inner queries $q_{2,4}$ and 
	no error to outer queries $q_{1,3,5}$.}}\label{scheme:compactor}
	\vspace*{-10pt}
\end{figure}

This intuition is implemented via a \emph{compactor}. First introduced by Manku~\etal~in~\cite{manku1998approximate}, 
it defines an array of $k$ items with weight $w$ each, and a compaction procedure which compress all $k$ items into $k/2$ items with weight $2w$. 
A compaction procedure first sorts all items, then deletes either even or odd positions and doubles the weight of the rest. 
Figure~\ref{scheme:compactor} depicts the error introduced for different rank queries $q$, by a compaction procedure applied to an example 
array of items $[1,3,5,8]$. 
Notice that the compactor utilizes the same idea as the one pair compression, but on the pairs of neighbors in the sorted array; thus by
performing $k/2$ non-intersecting compressions it  introduces an overall error of $w$ as opposed to $kw/2$.

The algorithm introduced in~\cite{manku1998approximate}, defines a stack of $H = O(\log{\frac{n}{k}})$ compactors, each of size $k$. 
Each compactor obtains as an input a stream and outputs a stream with half the size by performing a compact operation each time its buffer is full. 
The output of the final compactor is a stream of length $k$ that can simply be stored in memory. 
The bottom compactor that observes items has a weight of $1$; the next one observes items of weight $2$ and the top one $2^{H-1}$. 
The output of a compactor on $h$-th level is an input of the compactor on $(h+1)$-th level.
Note that the error introduced on $h$-th level is equal to the number of compactions $m_h = \frac{n}{kw_h}$ times the error introduced by one compaction $w_h$. 
The total error can be computed~as:~${\text{Err} = \sum_{h=1}^H{m_{h}w_h} = H\frac{n}{k} = O\left(\frac{n}{k} \log{\frac{n}{k}}\right)}.$ 
Setting~$k = O(\epss \log{\eps n})$ will lead to an approximation error of~$\eps n$.
The space used by $H$ compactors of size $k$ each is $O(\epss \log^2 {\eps n})$. 
Note that the algorithm is deterministic. 
Later, Agarwal~\etal~\cite{agarwal2013mergeable} suggested the compactor to choose the odd or even positions randomly and equiprobably, 
pushing the introduced error to zero in expectation. 
Additionally, the authors suggested a new way of feeding a subsampled streams into the data structure, 
recalling that $O(\frac{1}{\eps^2} \log \epss)$~samples preserve quantiles with $\pm \eps n$ approximation error. 
The proposed algorithm requires~$O(\epss \log^{3/2}{\epss})$ space and succeeds with high constant probability. 

To prove the result the authors introduced a random variable~$X_{i,h}$ denoting the error introduced on 
the~$i$-th compaction at $h$-th level. Then the overall error is
$\text{Err} = \sum_{h=1}^H{\sum_{i=1}^{m_h} {w_h X_{i,h}}},$
where~$w_h X_{i,h}$~is bounded, has mean zero and is independent of the other variables.
Thus, due to the Hoeffding's inequality:
\vspace*{-5pt}
$$P(|\text{Err}| > \eps n ) \le 2 \exp{\frac{-\eps^2n^2}{\sum_{h=1}^H{\sum_{i=1}^{m_h} {w_h^2}}}}.$$
\vspace*{-2pt}
Setting $w_h=2^{h-1}$ and $k =  O\left(\epss \sqrt{1/(\eps\delta)}\right)$ will keep the error probability 
bounded by $\delta$ for $O\left(\epss\right)$ quantiles. 

The following improvements were made by Karnin~\etal~\cite{karnin2016optimal}.
\begin{enumerate}
	\item Use exponentially decreasing size of the compactor. Higher weighted items receive higher capacity compactors.
	\item Replace compactors of capacity 2 with a sampler. This retains only the top $O(\log \epss)$ top compactors.
	\item Keep the size of the top $O(\log \log 1/\delta)$ compactors fixed.
	\item Replace the top $O(\log \log 1/\delta)$ compactors with a GK sketch~\cite{greenwald2001space}. 
\end{enumerate}
(1) and (2) reduced the space complexity to $O(\epss \sqrt{\log {1/\eps}})$, 
(3) pushed it further to $O(\epss \log^2 {\log{\epss}})$, and (4)
led to an optimal $O(\epss \log{\log{\epss}}).$ The authors also provided a matching lower bound. 
Note, the last solution is not mergeable due to the use of GK~\cite{greenwald2001space} as a subroutine. 

While (3) and (4) lead to the asymptotically better algorithm, its implementation is complicated for application purposes
and mostly are of a theoretical interest. 
In this paper we build upon the KLL algorithm of~\cite{karnin2016optimal} using only~(1)~and~(2).

In \cite{karnin2016optimal}, the authors suggest the size of the compactor to decrease as $k_h = c^{H-h}k$, for $c \in (0.5,1)$, 
then ${\sum_{h=1}^H\sum_{i=1}^{m_h}w_h^2 \le n^2/ (k^2C)}$ and  
$P(|\text{Err}|> \eps n) \le 2\exp{\left(-C\eps^2k^2\right)} \le \delta,$
where $C = 2c^2(2c-1)$. \footnote{In fact \cite{karnin2016optimal} has a fixable mistake in their derivation. 
For the sake of completeness in Appendix~\ref{app:kllfix} we clarify that the original results holds although with a slightly different constant terms.}
Setting $k = O\left(\epss\sqrt{\log{1/\eps}}\right)$ leads to the desired approximation guarantee for all $O\left(1/\eps\right)$ quantiles with constant probability. 
Note that the smallest meaningful compactor has size $2$, thus the algorithm will require 
$k(1 + c +\ldots  + c^{log_{1/c}{k}}) + O(\log n) = \frac{k}{1-c} + O(\log n)$  compactors,
where the last term is due to the stack of compactors of size $2$. The authors suggested replacing that stack with a basic sampler, 
which picks one item out of every $2^{w_{H - log_{1/c}{k}}}$ updates at random and logically is identical but consumes only $O(1)$ space. 
The resulting space complexity is $O(\epss\sqrt{\log{1/\eps}})$. We provide the pseudocode for the core routine in Algorithm~\ref{code:kll}.
\vspace*{-0.1in}
\begin{algorithm}
	\caption{Core routines for KLL algorithm \cite{karnin2016optimal} }\label{code:kll}
\begin{algorithmic}[1]
	\Function{KLL.update}{item}
	\State \algorithmicif\ \Call{Sampler}{item} \algorithmicthen\ \Call{KLL$[0]$.append}{item} 
	\For{$h = 1\ldots H$}
	\State \algorithmicif\ \Call{len}{KLL[$h$] $\ge$ $k_h$}  \algorithmicthen\ \Call{KLL.compact}{$h$} 
	\EndFor
	\EndFunction

	\Function{KLL.compact}{h}
	\State \Call{KLL[$h$].sort}{}(); rb = \Call{random}{\{0,1\}};
	\State \Call{KLL[$h+1$].extend}{}(KLL[$h$][rb : : $2$])
	\State KLL[$h$]= []
	\EndFunction
\end{algorithmic}
\end{algorithm}
\vspace*{-0.15in}

\section{Our Contribution}
Although the asymptotic optimum is already achieved for the quantile problem, there remains room for improvement from a practical perspective. In what follows we provide novel modifications to the existing algorithms that improve both their memory consumption and run-time. In addition to the performance, we ensure the algorithm is easy to use by (1) having the algorithm require only a memory limit, as opposed to versions that must know the values of $\eps, \delta$ in advance, and (2) by extending the functionality of the sketching algorithm to handle weighted examples.
We demonstrate the value of our algorithm in Section~\ref{section:experiments} with empirical experiments.
\subsection{Lazy compactions}
\label{sec:lazy}
\begin{figure}[t!]
	\centering
	\includegraphics[width=0.35\textwidth]{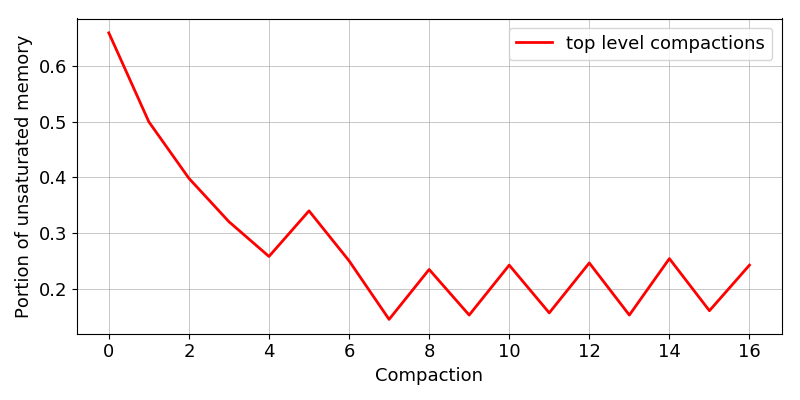}
	\vspace{-10pt}	
	\caption{Portion of unsaturated memory when compacting the top layer.}\label{fig:unsaturatedMemory} 
	\vspace{-10pt}	
\end{figure}	
Consider a simplified model, when the length of the stream $n$ is known in advance. One can easily identify the weight on the top layer 
of KLL data structure, as well as the sampling rate and the size of each compactor. Additionally, these parameters do not change while processing the stream.
Then note that while we are processing the first half of the stream, the top layer of KLL will be at most half full, 
i.e.~half of the top compactor memory will not be in 
use during processing first $n/2$ items.  
Let $s$ be the total amount of allocated memory and $c$ be the compactor size decrease rate. 
The top layer is of size $s(1-c)$, meaning that a fraction of $(1-c)/2$ is not used throughout that time period. 
The suggested value for $c$ is $1/\sqrt{2}$  which means that this quantity is $~15\%$. 
This is of course a lower estimate as the other layers in the algorithm are not utilized in various stages of the processing.
A similar problem arise when we do not know the final $n$ and keep updating it online: 
When the top layer is full the algorithm compacts it into a new layer;
at this moment the algorithm basically doubles its guess of the final~$n$. Although after this compaction $k/2$ items immediately appear on the top layer, 
we still have $1/4$ of the top layer not in use until the next update of $n$. This unused fraction accounts for $~7\%$ of the overall allocated memory.

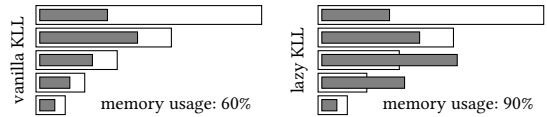
\begin{figure}[h!] 
	\begin{tikzpicture}[scale=1]

\draw (0, 0) rectangle (3, -0.25);
\draw[fill=gray] (0.05, -0.05) rectangle (0.95, -0.20);
\draw (0, -0.3) rectangle (3*0.6, -0.55);
\draw[fill=gray] (0.05, -0.35) rectangle (1.35, -0.50);
\draw (0, -.6) rectangle (3*0.6*0.6, -.85);
\draw[fill=gray] (0.05, -0.65) rectangle (0.75, -0.80);
\draw (0, -.9) rectangle (3*0.6*0.6*0.6, -1.15);
\draw[fill=gray] (0.05, -0.95) rectangle (0.45, -1.10);
\draw (0, -1.2) rectangle (3*0.6*0.6*0.6*0.6, -1.45);
\draw[fill=gray] (0.05, -1.25) rectangle (0.25, -1.40);
\node[right] at (0.75, -1.325) {\footnotesize memory usage: $60\%$};
\node[rotate=90] at(-0.25, -.725){\footnotesize vanilla KLL};

\draw (3.75 + 0, 0) rectangle (3.75 + 3, -0.25);
\draw[fill=gray] (3.75 + 0.05, -0.05) rectangle (3.75 + 0.95, -0.20);
\draw (3.75 + 0, -0.3) rectangle (3.75 + 3*0.6, -0.55);
\draw[fill=gray] (3.75 + 0.05, -0.35) rectangle (3.75 + 1.35, -0.50);
\draw (3.75 + 0, -.6) rectangle (3.75 + 3*0.6*0.6, -.85);
\draw[fill=gray] (3.75 + 0.05, -0.65) rectangle (3.75 + 1.85, -0.80);
\draw (3.75 + 0, -.9) rectangle (3.75 + 3*0.6*0.6*0.6, -1.15);
\draw[fill=gray] (3.75 + 0.05, -0.95) rectangle (3.75 + 1.15, -1.10);
\draw (3.75 + 0, -1.2) rectangle (3.75 + 3*0.6*0.6*0.6*0.6, -1.45);
\draw[fill=gray] (3.75 + 0.05, -1.25) rectangle (3.75 + 0.25, -1.40);
\node[right] at (3.75 + 0.75, -1.325) {\footnotesize memory usage: $90\%$};
\node[rotate=90] at(3.75 + -0.25, -.725){\footnotesize lazy KLL};

\end{tikzpicture}
	\caption{Compactor saturation: vanilla KLL vs.~lazy KLL} \label{scheme:lazy}
\end{figure}

We suggest all the compactors share the pool of allocated memory and perform a compaction only when the 
pool is fully saturated. This way each compaction is applied to a potentially larger set of items compared to the fixed budget setting, leading to less compactions.
Each compaction introduces a fixed amount of error thus the total error introduced is lower.  
Algorithm~\ref{code:lazy} gives the formal lazy-compacting algorithm,
and Figure~\ref{scheme:lazy} visualizes its advantage: in vanilla KLL all compactors having less items than their 
individual capacities, in lazy KLL this is not enforced due to sharing the pool of memory. In Figure~\ref{fig:unsaturatedMemory} you
can see that the memory is indeed unsaturated even when we compact the top level.

\begin{algorithm}[h!]
\caption{Update procedure for lazy KLL}\label{code:lazy}
\begin{algorithmic}[1]
		\Function{KLL.Update}{item}
		\State \algorithmicif\ \Call{Sampler}{item} \algorithmicthen\ \Call{KLL$[0]$.append}{item};\;itemsN++;     	\If{itemsN $>$ sketchSpace}
	    \For{$h = 1\ldots H$}
	    \If{\Call{len}{KLL[$h$]} $\ge$ $k_h$} 
	    \State \Call{KLL.compact}{$h$}; \textbf{break};
	    \EndIf
	    \EndFor
	\EndIf
	\EndFunction
\end{algorithmic}
\end{algorithm}

\subsection{Reduced randomness via Anti-Correlations} \label{sec:reduced_random}
Consider the process involving a single compactor layer. 
A convenient way of analyzing its performance is viewing it as a stream processing unit. 
It receives a stream of size $n$ and outputs a stream of size $n/2$. 
When collecting $k$ items it sorts them, and outputs (to the output stream) either those with even or odd locations. 
A deterministic compactor may admit an error of up to $n/k$. A compactor that decides whether to output the even or odds uniformly 
at random at every step admits an error of $\sqrt{n/k}$ in expectation as the directions of the errors are completely uncorrelated. 
Here we suggest a way to force a negative correlation that reduce the mean error by a factor of $\sqrt{2}$.
The idea is to group the compaction operations into pairs. 
At the $(2i)$-th compaction, choose uniformly at random whether to output the even or odd items, as described above.  
In the $(2i + 1)$-th compaction, perform the opposite decision compared to the $(2i)$-th compaction. 

This way, each coin flip defines $2$ consecutive compactions: 
with probability $\frac{1}{2}$ it is even $\rightarrow$ odd ($e\rightarrow o$), and with probability $\frac{1}{2}$ it is odd $\rightarrow$ even ($o\rightarrow e$).

Let's analyze the error under this strategy. Recall from Section~\ref{section:unifiedview} 
that for a rank query $q$ and a compaction operation, $q$ is either an inner or outer query. 
If it is an outer query, it suffers no error. If it is an inner query, it suffers and error of $+w$ if we output the odds 
and $-w$ if we output evens. Consider the error associated with a single query after two consecutive and anti-correlated compactions. 
We represent the four possibilities of $q$ as $io$(inner-outer), $oi$, $ii$, $oo$. 
\begin{table}[b!]
	\centering
	\begin{tabular}{c|c|c|c|c|c} 
		& $ii$ & $io$ & $oi$ & $oo$ &\\ 
		\hline
		$even\rightarrow odd$ & $0$ & $-w$ & $+w$ & $0$ & w.p. $1/2$ \\	
		\hline
		$odd\rightarrow even$ & $0$ & $+w$ & $-w$ & $0$ & w.p. $1/2$\\	
	\end{tabular}
	\caption{Error of a fixed rank query during two anti-correlated compactions}\label{table:oneflip}
\end{table}
Clearly, in expectation every two compactions introduce $0$ error. 
Additionally, we conclude that instead of suffering an error of up to $\pm w$ for every single compaction operation, 
we suffer that error for every two compaction operations. 
It follows that the variance of the error is twice smaller, hence the mean error is cut by a factor of $\sqrt{2}$.


\subsection{Error spreading} \label{sec:eq_spread}
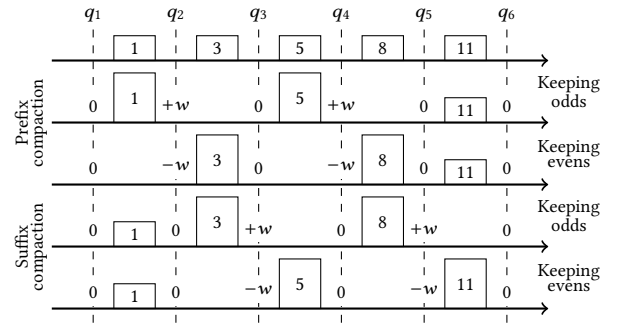
\begin{figure}[b!] 
	\centering
	\begin{tikzpicture}[scale=1.1]
\def\A{1}
\def\B{-0.75}

\baselineskip=2pt

\node[align=center] at (5.75,1.5*\B) {\footnotesize Keeping\\\footnotesize odds};
\node[align=center] at (5.75,2.5*\B) {\footnotesize Keeping\\\footnotesize evens};
\node[align=center,rotate=90 ] at (-0.75,2*\B) {\footnotesize Prefix \\\footnotesize compaction};
\node[align=center] at (5.75,3.5*\B) {\footnotesize Keeping\\\footnotesize odds};
\node[align=center] at (5.75,4.5*\B) {\footnotesize Keeping\\\footnotesize evens};
\node[align=center,rotate=90 ] at (-0.75,4*\B) {\footnotesize Suffix \\\footnotesize compaction};

\node[above ] at (0,0.5*\B) {\footnotesize $q_1$};
\draw[dashed] (0, 0.5*\B) -- (0,5.25*\B);
\node[fill=white, below ] at (0,1.5*\B) {\footnotesize $0$};
\node[fill=white, below ] at (0,2.5*\B) {\footnotesize $0$};
\node[fill=white, below ] at (0,3.5*\B) {\footnotesize $0$};
\node[fill=white, below ] at (0,4.5*\B) {\footnotesize $0$};

\node[above ] at (1,0.5*\B) {\footnotesize $q_2$};
\draw[dashed] (1, 0.5*\B) -- (1,5.25*\B);
\node[fill=white, below ] at (1,1.5*\B) {\footnotesize $+w$};
\node[fill=white, below ] at (1,2.5*\B) {\footnotesize $-w$};
\node[fill=white, below ] at (1,3.5*\B) {\footnotesize $0$};
\node[fill=white, below ] at (1,4.5*\B) {\footnotesize $0$};

\node[above ] at (2,0.5*\B) {\footnotesize $q_3$};
\draw[dashed] (2, 0.5*\B) -- (2,5.25*\B);
\node[fill=white, below ] at (2,1.5*\B) {\footnotesize $0$};
\node[fill=white, below ] at (2,2.5*\B) {\footnotesize $0$};
\node[fill=white, below ] at (2,3.5*\B) {\footnotesize $+w$};
\node[fill=white, below ] at (2,4.5*\B) {\footnotesize $-w$};

\node[above ] at (3,0.5*\B) {\footnotesize $q_4$};
\draw[dashed] (3, 0.5*\B) -- (3,5.25*\B);
\node[fill=white, below ] at (3,1.5*\B) {\footnotesize $+w$};
\node[fill=white, below ] at (3,2.5*\B) {\footnotesize $-w$};
\node[fill=white, below ] at (3,3.5*\B) {\footnotesize $0$};
\node[fill=white, below ] at (3,4.5*\B) {\footnotesize $0$};

\node[above ] at (4,0.5*\B) {\footnotesize $q_5$};
\draw[dashed] (4, 0.5*\B) -- (4,5.25*\B);
\node[fill=white, below ] at (4,1.5*\B) {\footnotesize $0$};
\node[fill=white, below ] at (4,2.5*\B) {\footnotesize $0$};
\node[fill=white, below ] at (4,3.5*\B) {\footnotesize $+w$};
\node[fill=white, below ] at (4,4.5*\B) {\footnotesize $-w$};

\node[above ] at (5,0.5*\B) {\footnotesize $q_6$};
\draw[dashed] (5, 0.5*\B) -- (5,5.25*\B);
\node[fill=white, below ] at (5,1.5*\B) {\footnotesize $0$};
\node[fill=white, below ] at (5,2.5*\B) {\footnotesize $0$};
\node[fill=white, below ] at (5,3.5*\B) {\footnotesize $0$};
\node[fill=white, below ] at (5,4.5*\B) {\footnotesize $0$};

\draw[->,thick] (-0.5, \B) -- (5.5,\B);
\draw[->,thick] (-0.5, 2*\B) -- (5.5, 2*\B);
\draw[->,thick] (-0.5, 3*\B) -- (5.5, 3*\B);
\draw[->,thick] (-0.5, 4*\B) -- (5.5, 4*\B);
\draw[->,thick] (-0.5, 5*\B) -- (5.5, 5*\B);

\draw (0.25,\B) rectangle (.75,\B+0.3)  node[pos=.5] {\footnotesize $1$} ;;
\draw (1.25,\B) rectangle (1.75,\B+0.3)  node[pos=.5] {\footnotesize $3$} ;;
\draw (2.25,\B) rectangle (2.75,\B+0.3)  node[pos=.5] {\footnotesize $5$} ;;
\draw (3.25,\B) rectangle (3.75,\B+0.3)  node[pos=.5] {\footnotesize $8$} ;;
\draw (4.25,\B) rectangle (4.75,\B+0.3)  node[pos=.5] {\footnotesize $11$} ;;

\draw (0.25,2*\B) rectangle (.75,2*\B+2*0.3)  node[pos=.5] {\footnotesize $1$} ;;
\draw (2.25,2*\B) rectangle (2.75,2*\B+2*0.3)  node[pos=.5] {\footnotesize $5$} ;;
\draw (4.25,2*\B) rectangle (4.75,2*\B+0.3)  node[pos=.5] {\footnotesize $11$} ;;

\draw (1.25,3*\B) rectangle (1.75,3*\B+2*0.3)  node[pos=.5] {\footnotesize $3$} ;;
\draw (3.25,3*\B) rectangle (3.75,3*\B+2*0.3)  node[pos=.5] {\footnotesize $8$} ;;
\draw (4.25,3*\B) rectangle (4.75,3*\B+0.3)  node[pos=.5] {\footnotesize $11$} ;;

\draw (0.25,4*\B) rectangle (0.75,4*\B+0.3)  node[pos=.5] {\footnotesize $1$} ;;
\draw (1.25,4*\B) rectangle (1.75,4*\B+2*0.3)  node[pos=.5] {\footnotesize $3$} ;;
\draw (3.25,4*\B) rectangle (3.75,4*\B+2*0.3)  node[pos=.5] {\footnotesize $8$} ;;

\draw (0.25,5*\B) rectangle (0.75,5*\B+0.3)  node[pos=.5] {\footnotesize $1$} ;;
\draw (2.25,5*\B) rectangle (2.75,5*\B+2*0.3)  node[pos=.5] {\footnotesize $5$} ;;
\draw (4.25,5*\B) rectangle (4.75,5*\B+2*0.3)  node[pos=.5] {\footnotesize $11$} ;;

\end{tikzpicture}
	\caption{Error analysis for a single query during a compaction. There are now four possibilities: prefix/suffix compaction, keep even/odd positions}\label{scheme:spread}
\end{figure}
Recall that in the analysis of all compactor based solutions~\cite{manku1998approximate,agarwal2013mergeable,karnin2016optimal,wang2013quantiles}. 
During a single compaction we can distinguish two types of rank queries: inner queries, for which some error is introduced, and outer queries, 
for which no error is introduced. 
Though the algorithms use this distinction in their analysis, they do not take an action to reduce the number of inner queries. 
It follows that for an arbitrary stream and an arbitrary query, the query may be an inner query the majority of the time, 
as it is treated in the analysis. In this section we provide a method that makes sure that a query has an 
equal chance of being inner or outer, thereby cutting in half the variance of the error associated with any query, 
for any stream. Consider a single compactor with a buffer of $k$ slots, and suppose $k$ is odd.  
On each compaction we flip a coin and then either compact the items with indices $1$ to $k-1$ (prefix compaction) 
or $2$ to $k$ (suffix compaction) equiprobably. This way each query is either inner or outer equiprobably. 
Formally, for a fixed rank query $q$:  with probability at least $\frac{1}{2}$ it is an outer query and then no error is introduced,
with probability at most $\frac{1}{4}$ it is an inner query with error $-w$; and with probability at most $\frac{1}{4}$ it is an inner 
with error $+w$. We thus still have an unbiased estimator for the query's rank but the variance is cut in half. 
We note that the same analysis applies for two consecutive compactions using the reduced randomness 
improvement discussed in Section~\ref{sec:reduced_random}: The configuration ($ii$,$io$,$oi$,$oo$) of a query in two consecutive 
compactions described in Table~\ref{table:oneflip} will now happen with equal probability, 
hence we have the same distribution for the error: $0$ with probability at least $\frac{1}{2}$, $+w$ and $-w$ with probability at most $\frac{1}{4}$ each,
meaning that the variance is cut in half compared to its worse case analysis without the error-spreading improvement. 
Figure~\ref{scheme:spread} visualizes the analysis of the error for a fixed query during a single compaction operation.	

\subsection{Sweep-compactor}
\label{sec:sweep}
The error bound for all compactor based algorithms follows from the property that every batch of $k/2$ pair compressions is disjoint. 
In other words, the compactor makes sure that all of the compacted pairs can be partitioned into sets of size exactly $k/2$, 
the intervals corresponding to each set are disjoint, and the error bound is a result of this property.
In this section we provide a modified compactor that compacts pairs one at time while maintaining the guarantee 
that pairs can be split into sets of size \emph{at least} $k/2$ such that the intervals corresponding to the pairs 
of each set are disjoint.
Compacting a single pair takes constant time; hence we reduce the worst-case update time from $O(\epss)$ to $O(\log\epss)$. 
Additionally, for some data streams the disjoint batch size is strictily larger than $k/2$ resulting in a reduction in the overall error. 

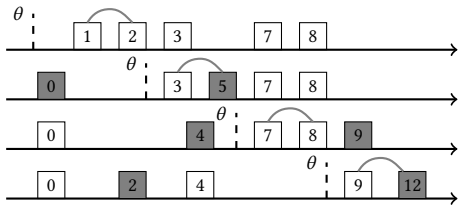
\begin{figure}[!b] 
	\begin{tikzpicture}[scale=1.2]

\def\B{-0.55}
\def\C{0.3}

\draw[->,thick] (-0.5, 1*\B) -- (4.5,1*\B);
\draw[->,thick] (-0.5, 2*\B) -- (4.5,2*\B);
\draw[->,thick] (-0.5, 3*\B) -- (4.5,3*\B);
\draw[->,thick] (-0.5, 4*\B) -- (4.5,4*\B);

\draw (0.25,1*\B) rectangle (.55, .3 + 1*\B) node[pos=.5] {\footnotesize $1$} ;
\draw (0.75,1*\B) rectangle (1.05,.3 + 1*\B) node[pos=.5] {\footnotesize $2$} ;
\draw (1.25,1*\B) rectangle (1.55,.3 + 1*\B) node[pos=.5] {\footnotesize $3$} ;
\draw (2.25,1*\B) rectangle (2.55,.3 + 1*\B) node[pos=.5] {\footnotesize $7$} ;
\draw (2.75,1*\B) rectangle (3.05,.3 + 1*\B) node[pos=.5] {\footnotesize $8$} ;

\draw [thick, gray] (0.4 ,0.3 + 1*\B) to [out=60, in=180](0.4/2 + 0.9/2, 0.3 + 1*\B + 0.15) to [out=0,in=135] (0.9 ,0.3 + 1*\B );  

\draw[dashed,thick] (-0.2,  + 1*\B) -- (-0.2,0.4 + 1*\B) node[left] {\footnotesize $\theta$};
\draw [fill=gray](-0.15,2*\B) rectangle (.15, .3 + 2*\B) node[pos=.5] {\footnotesize $0$} ;
\draw (1.25,2*\B) rectangle (1.55,.3 + 2*\B) node[pos=.5] {\footnotesize $3$} ;
\draw [fill=gray](1.75,2*\B) rectangle (2.05,.3 + 2*\B) node[pos=.5] {\footnotesize $5$} ;
\draw (2.25,2*\B) rectangle (2.55,.3 + 2*\B) node[pos=.5] {\footnotesize $7$} ;
\draw (2.75,2*\B) rectangle (3.05,.3 + 2*\B) node[pos=.5] {\footnotesize $8$} ;

\draw [thick, gray] (1.4 ,0.3 + 2*\B) to [out=60, in=180](1.4/2 + 1.9/2, 0.3 + 2*\B + 0.15) to [out=0,in=135] (1.9 ,0.3 + 2*\B );  

\draw[dashed,thick] (1.05,  + 2*\B) -- (1.05,0.4 + 2*\B) node[left] {\footnotesize $\theta$};
\draw (-0.15,3*\B) rectangle (.15, .3 + 3*\B) node[pos=.5] {\footnotesize $0$} ;
\draw [fill=gray](1.5,3*\B) rectangle (1.8,.3 + 3*\B) node[pos=.5] {\footnotesize $4$} ;
\draw (2.25,3*\B) rectangle (2.55,.3 + 3*\B) node[pos=.5] {\footnotesize $7$} ;
\draw (2.75,3*\B) rectangle (3.05,.3 + 3*\B) node[pos=.5] {\footnotesize $8$} ;
\draw [fill=gray](3.25,3*\B) rectangle (3.55,.3 + 3*\B) node[pos=.5] {\footnotesize $9$} ;

\draw [thick, gray] (2.4 ,0.3 + 3*\B) to [out=60, in=180](2.4/2 + 2.9/2, 0.3 + 3*\B + 0.15) to [out=0,in=135] (2.9 ,0.3 + 3*\B );  

\draw[dashed,thick] (2.05,  + 3*\B) -- (2.05,0.4 + 3*\B) node[left] {\footnotesize $\theta$};
\draw (-0.15,4*\B) rectangle (.15, .3 + 4*\B) node[pos=.5] {\footnotesize $0$} ;
\draw (0.75,4*\B)[fill=gray] rectangle (1.05,.3 + 4*\B) node[pos=.5] {\footnotesize $2$} ;
\draw (1.5,4*\B) rectangle (1.8,.3 + 4*\B) node[pos=.5] {\footnotesize $4$} ;
\draw (3.25,4*\B) rectangle (3.55,.3 + 4*\B) node[pos=.5] {\footnotesize $9$} ;
\draw [fill=gray](3.85,4*\B) rectangle (4.15,.3 + 4*\B) node[pos=.5] {\footnotesize $12$} ;

\draw [thick, gray] (3.4 ,0.3 + 4*\B) to [out=60, in=180](3.4/2 + 3.9/2, 0.3 + 4*\B + 0.15) to [out=0,in=135] (3.9 ,0.3 + 4*\B );  
\draw[dashed,thick] (3.05,  + 4*\B) -- (3.05,0.4 + 4*\B) node[left] {\footnotesize $\theta$};
%
%
%
%
\end{tikzpicture} 
	\caption{The inner state of a sweep-compactor during a single sweep operation. 
	Notice that in this example, although we have a buffer of size $k=5$, a single sweep managed to compress $4$ pairs, 
	rather than $\lfloor k/2 \rfloor = 2$}\label{scheme:sweep}
\end{figure}

The modified compactor operates in phases we call \emph{sweeps}. 
It maintains the same buffer as before and an additional threshold $\theta$ initialized as special \texttt{null} value. 
The items in the buffer are stored in non-decreasing sorted order. When we reach capacity we compact a single pair. 
If $\theta$ is \texttt{null} we set it to $-\infty$\footnote{Notice that $-\infty$ is still defined in the comparison model.}
or to the value of the smallest item uniformly at random. 
This mimics the behavior of the prefix/suffix compressions of Section~\ref{sec:eq_spread}
\footnote{If we wish to ignore prefix/suffix compactions $\theta$ should always be initialized to $-\infty$.}.
The pair we compact is a pair of consecutive items where the smaller item is the smallest item in the buffer that 
is larger than $\theta$.\footnote{We ignore the case of items with equal value. Note that if that happens, 
these two items should be compacted together as this is guaranteed not to incur a loss.}
If no such pair exist due to $\theta$ being too large, we start a new sweep, 
meaning we set $\theta$ to \texttt{null} and act as detailed above.
We note that a sweep is the equivalent to a compaction of a standard compactor. 
Due to this reason, we consistently keep either the smaller or larger item when compacting a single pair throughout a sweep. 
To keep true to the technique of Section~\ref{sec:reduced_random} we have sweep number $2i+1$ 
draw a coin to determine if the small or large items are kept, and sweep number $2i+2$ does the opposite. 
The pseudo-code for the sweep-compactor is given in Algorithm~\ref{code:sweep} and 
Figure~\ref{scheme:sweep} visualizes the inner state of the sweep-compactor during a single sweep.

%
%
%
%
\begin{algorithm}
	\caption{Sweep compaction procedure}\label{code:sweep}
\begin{algorithmic}[1]
	\Function{KLLsweep.compact}{h}
	\State \Call{KLL[$h$].sort}{}()
	\State $i^* = \argmin_i{(\text{KLL}[h][i] \ge \text{KLL}[h].\theta)}$
	\State \algorithmicif\ $i^* == $ \textbf{None} \algorithmicthen\ $i^* = 0$;  
	\State KLL$[h].\theta =$ KLL$[i^* + 1]$;
	\State \Call{KLL[$h$].pop}{}($i^* + $ \Call{randBit}{}());
	\State \textbf{return} \Call{KLL[$h$].pop}{}($i^*$)
	\EndFunction
\end{algorithmic}
\end{algorithm}
Notice that for an already sorted stream the modified compactor performs only a single sweep, 
hence in this scenario the resulting error would not be a sum of $n/k$ i.i.d.\ error terms, each of magnitude $\pm w$ 
but rather a single error term of magnitude $\pm w$. Though this extreme situation may not happen very often, 
it is likely that the data admits some sorted subsequences and the average sweep would contain more than $k/2$ pairs. 
We demonstrate this empirically in our experiments.

%
%
%
%
%
%
%
%

\section{Weighted stream extension} 
In the current section, we extend the existing quantiles algorithms to handle weighted inputs. 
Consider the stream of updates $(a_i, w_i)$, where each item $a_i$ comes with a weight $w_i$. 
After feeding the stream into a data structure, it is queried with a quantile $\phi$, and should report an item $x$ such that: 
$$(\phi - \eps) W \le \sum_{i: a_i < x}{w_i} \le  (\phi + \eps) W, $$
where $W = \sum_i{w_i}$ is the total weight of the entire stream.

Such a scenario may come up if the observed samples have different importance, e.g.,\ in the case of boosted decision trees where 
examples are re-weighted according to the current loss corresponding to them \cite{chen2016xgboost}, or load balancing of tasks with 
a different associated cost. It can also occur when we are observing samples from a 
set that were not chosen uniformly at random, or when we suffer a distribution drift and wish to give preference to more recent items.

One can approach the problem naively: break down each update 
$(a_i,w_i)$ into $w_i$ unitary updates and feed it into lazy sweeping KLL. 
In the worst case, the time to process one unitary update is $O(\log \epss)$, 
and the time to process one weighted update is~${O(\max(w_i)\log\epss)}$.

However, in a common scenario, weights $w_i$ do not increase exponentially with $i$. 
In this case for long enough streams, vast majority of updates $(a_i,w_i)$  
would satisfy $w_i \ll W(i)$, and in particular $w_i < 2^{H_s}$,
where $2^{H_s}$ is the sampling rate of the KLL sampler object. 

Recall, that in KLL the sampler maintains a reservoir sample of a single item until 
it observes $2^{H_s}$ items and outputs the sample to the stream observed by the first compactor. 
Then compactor processes their input stream and provides an output stream to the next compactor and so on.

In \cite{karnin2016optimal}, in order to obtain mergeable sketches the sampler 
object is in fact defined in a way that it can accept weighted inputs. 
It feeds the inputs into a weighted reservoir sample until that weight is larger than $2^{H_s}$. 
At that point, the sampler has the reservoir sample of weight $w_1$ and a new item of weight $w_2$. 
One of these items is being outputted into the bottom compactor input stream with 
a weight of exactly $2^h$, with probabilities that ensure an unbiased error. 
Weighted reservoir sampler can process updates with the weights less than the sampling rate 
in $O(1)$ time. Therefore, if $w_i$ does not grow exponentially, then in the worst-case the 
update time for the majority of updates becomes $O(\log \epss)$.


Further, we provide two approaches for handling the weighted input scenario in the general case, 
where we do not assume the slow growth of $w_i$.
The first is achieved via a near black-box approach, wrapping the KLL algorithm 
and manipulating the input data, which introduces extra $O(\epss\sqrt{\log\epss})$ overhead to the 
worst-case update time and $O(\log \epss)$ overhead to the 
amortized update time. In the second algorithm, we modify the core component, 
the compactor. It obtains a compactor that can handle items of different weight and uses the KLL 
paradigm with these new compactors to handle weighted inputs. 
The second approach does not suffer from the overhead of manipulating the incoming data 
and offers the same asymptotic run-time as the unweighted version.

\subsection{Splitting the Input}
The first algorithm we provide is obtained via a black box approach on the top of KLL for unweighted streams. 
Let $H$ be the minimal integer such that the total observed weight $W$, including that of a newly observed 
item has the property $W < k2^H$, where $k$ is the size of the top compactor,
and $2^{H_s}$ is the sampling rate.  
When a new items of weight $w$ is fed to the stream we view it in a 
(partially) binary representation: 

{\centering$w = w' + \sum_{h=H_s}^H a_h 2^h$\par}

\begin{figure}[t!] 
	\begin{tikzpicture}[scale=1.2]
\draw (0-1, 0) rectangle 	(-1 + 3, -0.25);
\draw (0-1, -0.3) rectangle 	(-1 + 3*0.6, -0.55);
\draw (0-1, -.6) rectangle 	(-1 + 3*0.6*0.6, -.85);
\draw (0-1, -.9) rectangle 	(-1 + 3*0.6*0.6*0.6, -1.15);
\draw (0-1, -1.2) rectangle 	(-1 + 3*0.6*0.6*0.6*0.6, -1.45);
\draw (0-1, -1.5) rectangle 	(-1 + 3*0.6*0.6*0.6*0.6*0.6, -1.75);
\node at (-1 + 3*0.6*0.6/2, -.125) {\tiny $w_H = 2^8$}; 
\node[right] at (-1 + 0.2, -1.65) {\tiny Sampler: $H_s=3$}; 
\node at (-4.25, +0.125 -1.2) {\footnotesize item $(a,861) = $};

\draw [dotted] (-2.9, -0.125 -0) to [out=245, in=115](-2.9, -0.125 -2.4);	
\node at (-2.5, -0.125 -0)  {\footnotesize$2^8\cdot 3$};
\draw[dotted] (-2.1, -0.125 -0) -- (-1.05, -0.125 -0);
\node[above] at (-1.60, -0.2 -0) {\tiny $3\times\;$push $a$};
\node at (-2.5, -0.125 -0.3){\footnotesize$2^7\cdot 0$};
\node at (-2.5, -0.125 -.6) {\footnotesize$2^6\cdot 1$};
\draw[dotted] (-2.1, -0.125 -.6) -- (-1.05, -0.125 -.6);
\node[above] at (-1.5, -0.2 -.6) {\tiny push $a$};
\node at (-2.5, -0.125 -.9) {\footnotesize$2^5\cdot 0$};
\node at (-2.5, -0.125 -1.2){\footnotesize$2^4\cdot 1$};
\draw[dotted] (-2.1, -0.125 -1.2) -- (-1.05, -0.125 -1.2); 
\node at (-1.5, -0.125 -1.125){\tiny push $a$};
\node at (-2.5, -0.125 -1.5){\footnotesize$2^3\cdot 1$};
\node at (-2.5, -0.125 -1.8){\footnotesize$2^2\cdot 1$};
\node at (-2.5, -0.125 -2.1){\footnotesize$2^1\cdot 0$};
\node at (-2.5, -0.125 -2.4){\footnotesize$2^0\cdot 1$};
\draw [dotted] (-2.1, -0.125 -1.5) to [out=295, in=65](-2.1, -0.125 -2.4);	
\draw[dotted] (-1.93, -0.125 -1.95) -- (-1.05, -0.125 -1.5); 
\node[] at (-1.5, -0.125 -1.95){\tiny push};
\node[] at (-1.5, -0.325 -1.95){\tiny$(a,13)$};


\end{tikzpicture}
	\vspace{-10pt}	
	\caption{Intuition behind base2update algorithm}\label{scheme:weightednaive2}
	\vspace{-10pt}	
\end{figure}
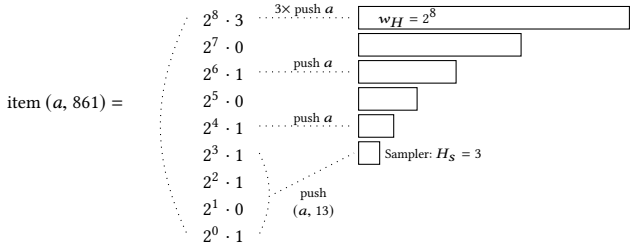

Here, $w' < 2^{H_s}$; for $h < H$, $a_h$ are either 0 or 1, and for $h=H$, $a_h$ can take any integer value between 0 and $k-1$. 
We feed the item with weight $w'$ to the sampler, then for all $h<H$ for which $a_h=1$ we feed the item to the appropriate compactor of level $h$.
Finally, we feed $a_H$ copies of the item to the compactor at the top~level~$H$. 
The process is depicted on the Figure~\ref{scheme:weightednaive2}. 

The compute has one additional step in case $H_{s}$ increased due to the new item. 
In this case, all items from the compactors  outputting items of weight $\leq 2^{H_s}$ are fed to the sampler. 
The entire process is given in Algorithm~\ref{code:base2update}. It implements the update function adding an item $a$ with a weight $w$ to the sketch. 
It invokes in a black box manner the sub-procedure we call KLL.pushItems that implements the item update of the unweighted KLL algorithm. 
This sub-procedure can either add an arbitrarily weight value for an item that is added to the sampler, 
or a weight equal to a power of two, determining the compactor that will receive the item.
\vspace{-5pt}
\begin{algorithm}
	\caption{Base2update update procedure}\label{code:base2update}
\begin{algorithmic}[1]
\small
	\Function{update}{$(a,w)$}
		\State  $H = \argmin_h(k2^h > W)$
		\State $H_{\text{new}} = \argmin_h (k2^h > W+w)$
		\If{$H_{\text{new}} > H$}\label{codelines:base2update:1}
		        \State delete bottom $H_{\text{new}} - H$ compactors 
			\State \Call{KLL.pushItems}{~all items from the deleted compactors~} \label{codelines:base2update:12}
			\State add $H_{\text{new}} - H$ empty compactors on the top; $H = H_{\text{new}}$;\label{codelines:base2update:2}
		\EndIf 
		\State let $2^{H_s}$ be the current sample rate 
		\State decompose $w = w' + \sum_{h=H_s}^H a_h2^h$ 
		\State \Comment{$\forall h<H:a_h\in\{0,1\};\; a_H\in[k];\; w' \le 2^{H_s}$}
		\State \Call{KLL.pushItems}{$a$, $w'$} 				\label{codelines:base2update:3}	
		\For{$h < H$ s.t. $a_h \neq 0$} \Call{KLL.pushItems}{$a$, $2^h$} \label{codelines:base2update:4} 
		\EndFor
		\State $a_H$ times \textbf{repeat}: \Call{KLL.pushItems}{$a$, $2^H$}\label{codelines:base2update:5}
	\EndFunction
\end{algorithmic}
\end{algorithm}
\vspace{-10pt}
\normalsize
\begin{theorem} \label{thm:base2update}
Algorithm~\ref{code:base2update} processes a stream of weighted updates and outputs all $\eps$-approximate quantiles with high probability using
memory $O(\epss\sqrt{\log1/\eps})$. In the worst-case scenario, a single update invokes $O(\log1/\eps)$ update calls to compactors of the KLL sketch, 
and $O(\epss\sqrt{\log1/\eps})$ calls to the sampler, resulting in a $O(\epss\log^{3/2}\epss)$ worst-case run-time. 
The amortized run-time is $O(\log^2\epss)$.
\end{theorem}

Due to space restrictions we defer the proof to Appendix~\ref{app:missing proofs}.

\subsection{Weight-aware Compactor}
Here we suggest a solution which does not require any stream transformation. 
Instead, we modify the main building block, the compactor, to handle different weights for its inputs. 
We define a weight aware compactor as an object that receives a stream of items of weights in $[w, 2w)$ for some scalar $w$ and 
outputs items of weight $[2w, 4w)$.

Suppose you are given two pairs $(a, w_a)$ and $(b, w_b)$, such that $a < b$, however, the data structure can store only one item. 
Due to the limitations of the comparison model, as described in section~\ref{section:unifiedview}, the only option is to pick either $a$ or $b$. 
The weight-aware compactor chooses $a$ with probability $\frac{w_a}{w_a + w_b}$ and $b$ with probability $\frac{w_b}{w_a+w_b}$, 
assigns weight $w_a + w_b$ for the chosen item and drop the other one. 

\begin{figure}[t!] 
	\begin{tikzpicture}[scale=1.5]

\coordinate (A) at (2,3);

\pgfmathsetmacro{\ax}{0}
\pgfmathsetmacro{\ay}{-1}
\pgfmathsetmacro{\toplen}{4}
\pgfmathsetmacro{\bottomlen}{2}
\pgfmathsetmacro{\wheight}{0.20}
\pgfmathsetmacro{\bx}{-0.5}
\pgfmathsetmacro{\by}{-2.1}
\pgfmathsetmacro{\cx}{2.5}
\pgfmathsetmacro{\cy}{-2.1}

\draw[dotted] (\ax,\ay + \wheight*2) node[left]{\footnotesize $2w$}-- (\ax + \toplen,\ay +\wheight*2);
\draw[dotted] (\ax,\ay + \wheight) node[left]{\footnotesize $w$}-- (\ax + \toplen,\ay +\wheight);
\draw[->,thick] (\ax,\ay) -- (\ax + \toplen,\ay);

\draw[fill=white] (\ax + 0.1*\toplen,\ay) rectangle (\ax + 0.15*\toplen,\ay + 1.1 * \wheight)  node[pos =0.5] {\footnotesize a};
\draw[fill=white] (\ax + 0.2*\toplen,\ay) rectangle (\ax + 0.25*\toplen,\ay + 1.2 * \wheight) node[pos=0.5]{\footnotesize b};
\draw[fill=white] (\ax + 0.4*\toplen,\ay) rectangle (\ax + 0.45*\toplen,\ay + 1.3 * \wheight) node[pos=0.5]{\footnotesize c};
\draw[fill=white] (\ax + 0.6*\toplen,\ay) rectangle (\ax + 0.65*\toplen,\ay + 1.9 * \wheight) node[pos=0.5]{\footnotesize d};
\draw[fill=white] (\ax + 0.8*\toplen,\ay) rectangle (\ax + 0.85*\toplen,\ay + 1.0 * \wheight) node[pos=0.5]{\footnotesize e};
\draw[fill=white] (\ax + 0.9*\toplen,\ay) rectangle (\ax + 0.95*\toplen,\ay + 1.5 * \wheight) node[pos=0.5]{\footnotesize f};

\draw[dotted] (\bx,\by + \wheight*4) node[left] {\footnotesize $4w$} -- (\bx + \bottomlen,\by +\wheight*4);
\draw[dotted] (\bx,\by + \wheight*2) node[left] {\footnotesize $2w$} -- (\bx + \bottomlen,\by +\wheight*2);
\draw[->,thick] (\bx,\by) -- (\bx + \bottomlen,\by);

\draw[fill=white] (\bx + 0.3*\bottomlen,\by) rectangle (\bx + 0.4*\bottomlen,\by + 3.2 * \wheight) node[pos=0.5] {\footnotesize c};


\draw[dotted] (\cx,\cy + \wheight*4) -- (\cx + \bottomlen,\cy +\wheight*4) node[right] {\footnotesize $4w$} ;
\draw[dotted] (\cx,\cy + \wheight*2) -- (\cx + \bottomlen,\cy +\wheight*2) node[right] {\footnotesize $2w$};
\draw[->,thick] (\cx,\cy) -- (\cx + \bottomlen,\cy);

\draw[fill=white] (\cx + 0.6*\bottomlen,\cy) rectangle (\cx + 0.7*\bottomlen,\cy + 3.2 * \wheight) node[pos=0.5] {\footnotesize d};

\draw[dashed] (\ax + 0.35*\toplen,\ay-0.2) rectangle (\ax + 0.7*\toplen,\ay + 1.3 * \wheight + 0.2);
\node[above left, fill=white] at (\bx + 1.0*\bottomlen,\by + 1 * \wheight) {\footnotesize w.p. $\frac{w_c}{w_c+w_d}$};
\node[above right, fill=white] at (\cx - 0*\bottomlen,\cy + 1*\wheight){\footnotesize w.p. $\frac{w_d}{w_c+w_d}$};;

\end{tikzpicture}
	\caption{Compressing pair in the weighted compactor}\label{scheme:weightednonnaive}
	\vspace{-10pt}
\end{figure}
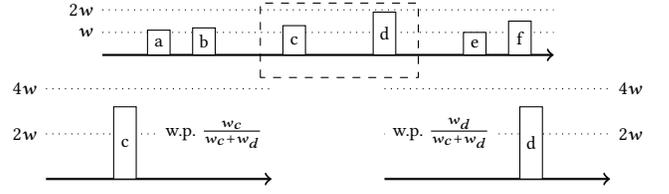

To carefully control the variance we define the weight-aware compactor as an array of pairs $\{(a_i, w_i)\}_{i=1}^{k_h}$ 
such that $w_i \in [w, 2w)$ for some pre-defined scalar $w$, and the compaction procedure is similar to the unweighted case: 
\begin{enumerate}
	\item sort the array using $a_i$ as an index
	\item break the array into pairs of neighbors ${(a_j, w_j), (a_{j+1}, w_{j+1})}$
	\item compress each pair, using procedure described above
\end{enumerate}
The intuition behind the weighted pair compression is depicted in the Figure~\ref{scheme:weightednonnaive} and the rest of process
is given in Algorithm \ref{code:base2compactor}, that due to space restrictions is available in the Appendix. 
\vspace{-5pt}
\normalsize
\begin{lemma}\label{lemma:base2compactor} 
Given a stream of $n$ items of weights in $[w,2w)$, a weight-aware compactor outputs a stream of $n/2$ items of weight $[2w,4w)$. If the memory budget of the weight-aware compactor is $k$ we have that for any query $q$, the error in its rank in the output stream compared to the input stream is equal to $\sum_{i=1}^{n/k} X_i$. Here, the $X_i$'s are independent random variables. For every $X_i$ we have $\text{E}[X_i]=0$ and $|X_i| < 2w$ w.p.\ 1.
\end{lemma}
Due to space restrictions we defer the proof to Appendix~\ref{app:missing proofs}.

The new algorithm will operate as the unweighted version of KLL did. It maintains a hierarchy of compactors, and a sampler at the bottom hierarchy. 
A compactor at level $h$ accepts inputs of weights in $[2^h, 2^{h+1})$, instead of exactly $2^h$ as in the unweighted case. 
As before, the sampler outputs items of weight $2^{H_s}$ and accepts items of weight in range from $1$ to $2^{H_s}$.
\begin{theorem} \label{thm:base2compactor}
	Algorithm \ref{code:base2compactor} processes a stream of weighted updates and outputs all $\eps$-approximate 
	quantiles with high probability using space $O(\epss\sqrt{\log1/\eps})$ and has both worst-case runtime and amortized runtime equal $O(\log1/\eps)$. 
\end{theorem}
Due to space restrictions we defer the proof to Appendix~\ref{app:missing proofs}.

%
%
%
%
%
%
%
%
%
%
%
%

\section{Experimental Results} \label{section:experiments}
\subsection{Data Sets}
To study the algorithms properties we tested it on both synthetic and real datasets, with various sizes, underlying distributions and orders.
Note that all the approximation guarantees of the investigated algorithms do not depend on the order in the data, 
however in practice the order might significantly influence the precision of the output within the theoretical guarantees. 
Surprisingly the worst-case is achieved when the dataset is randomly shuffled. Therefore, we will pay more attention to  
randomly ordered data sets in this section. We also experiment with the semi-random orders that resemble more to real life applications. 
Due to the space limitations we could not possibly present all the experiments in the paper and present here only the most interesting findings. 

Our experiments were carried on following synthetic datasets.\\
\textbf{Sorted} is a stream with all unique items in ascending order.\\
\textbf{Shuffled} is a randomly shuffled stream with all unique items.\\
\textbf{Trending} is $s_t = t/n\; + $ mean-zero random variable. Trending stream mimics a statistical drift over time 
(widely used in ML).\\ 
\textbf{Brownian} simulates a Brownian motion or a random walk which generates time series data not unlike CPU usage, stock market, traffic congestion, etc.
 The length of the stream varies from $10^5$ to $10^9$ for all the datasets.
%
%

In addition to synthetic data we use two publicly available datasets.
The first contains text information and the second contains IP addresses. Both objects types have a natural order and can be fed as input to a quantile sketching algorithm.

\noindent {\bf (1) Anonymized Internet Traces 2015 (CAIDA) \cite{caida}}
The dataset contains anonymized passive traffic traces from the internet data collection monitor which belongs to CAIDA 
(Center for Applied Internet Data Analysis) and
located at an Equinix data center in Chicago, IL. 
For simplicity we work with the stream of pairs $(\text{IP}_{\text{source}},\text{IP}_{\text{destination}})$. The comparison model is lexicographic. 
We evaluate the performance on the prefixes of the dataset of different sizes: from $10^7$ to $10^9$. Note that evaluation of the CDF of the underlying 
distribution for traffic flows helps optimize packet managing. CAIDA's datasets are used widely for verifying 
different sketching techniques to maintain different statistics over the flow, and finding quantiles and heavy hitters specifically.  

\noindent {\bf (2) Page view statistics for Wikimedia projects (Wiki) \cite{wiki}}
The dataset contains counts for the number of requests for each page of the Wikipedia project during 8 months of 2016. The data is aggregated by 
day, i.e. within each day data is sorted and each item is assigned with a count of requests during that day. Every update in this dataset is the
title of a Wikipedia page. We will experiment with both the original dataset and with its shuffled version.
Similarly to CAIDA we will consider for the Wiki dataset prefixes of size from $10^7$ to $10^9$. 
In our experiments, each update is a string containing the name of the page in Wikipedia. The comparison model is lexicographic.  

\begin{figure*}[h!]
	\centering
	\begin{subfigure}[]{0.33\textwidth}
		\includegraphics[width= \textwidth]{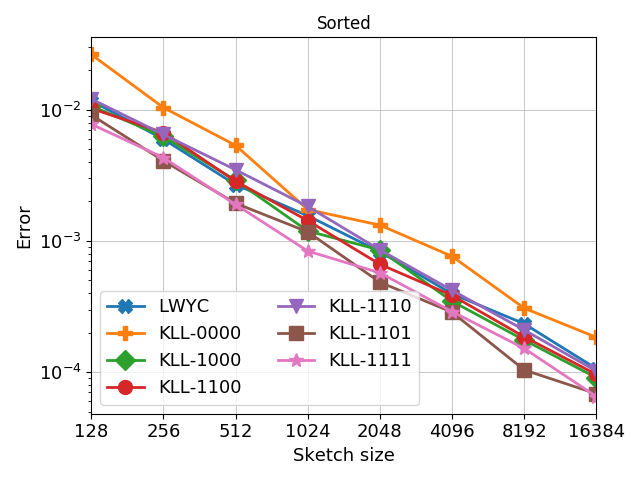}
		\caption{}\label{fig:sorted} 
	\end{subfigure}%
	\begin{subfigure}[]{0.33\textwidth}
		\includegraphics[width= \textwidth]{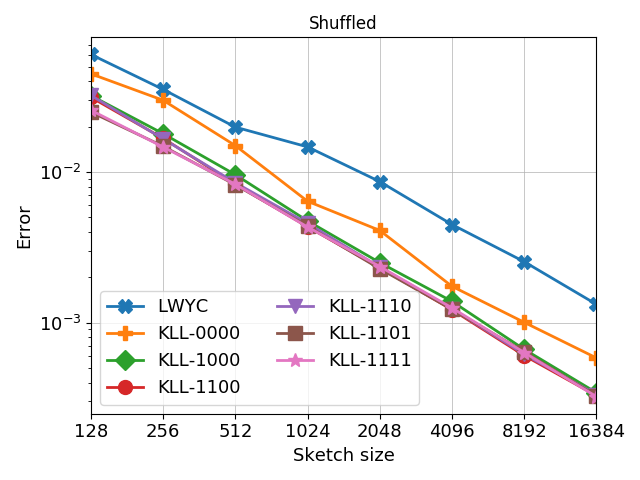}
		\caption{}\label{fig:shuffle}
	\end{subfigure}	
	\begin{subfigure}[]{0.33\textwidth}
		\includegraphics[width= \textwidth]{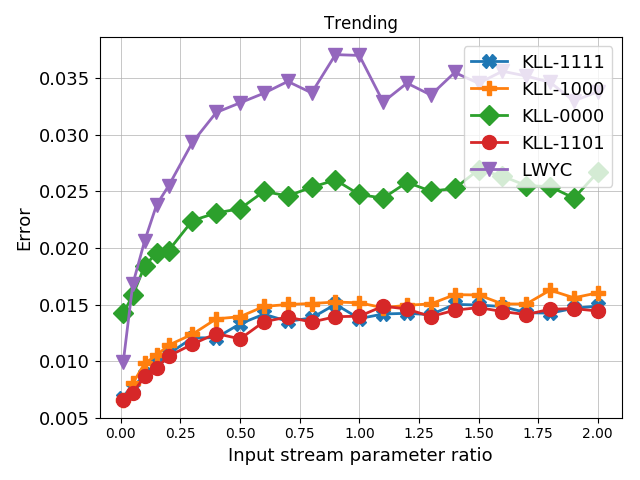}
		\caption{}\label{fig:trending}
	\end{subfigure}	
	\begin{subfigure}[]{0.33\textwidth}
		\includegraphics[width= \textwidth]{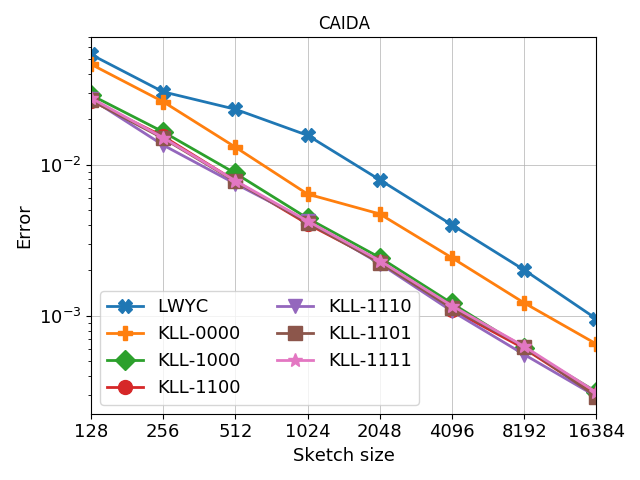}
		\caption{}\label{fig:caida}
	\end{subfigure}%
	\begin{subfigure}[]
	{0.33\textwidth}
		\includegraphics[width= \textwidth]{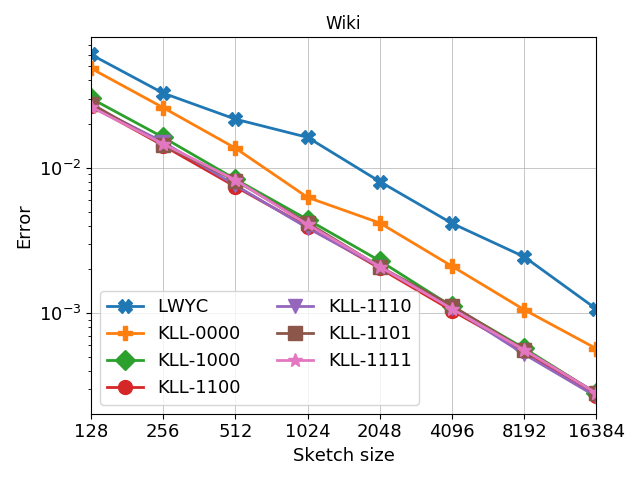}
		\caption{}\label{fig:wiki}
	\end{subfigure}
	\begin{subfigure}[]{0.33\textwidth}
		\includegraphics[width= \textwidth]{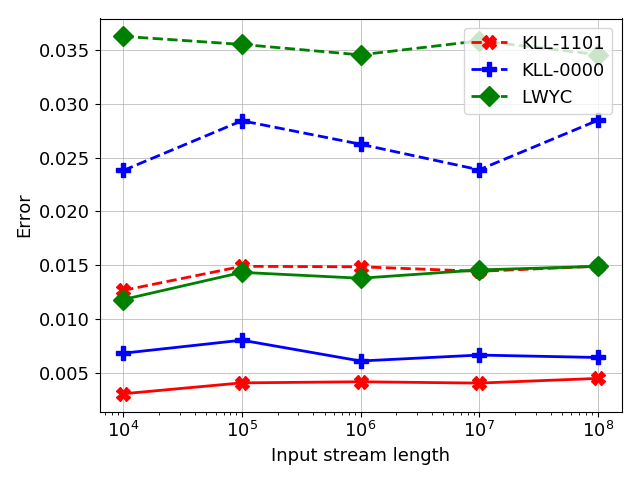}
		\caption{}\label{fig:constant}
	\end{subfigure}	
	
	\begin{subfigure}[]{0.45\textwidth}
		{\small 
		\begin{tabular}{|l||*{5}{c|}}\hline
		\diagbox{ sketch}{ size}&$  2^7$&$2^8$&$2^9$&$2^{10}$&$2^{11}$ \\\hline\hline
				 	LWYC  	 &0.0602&0.0352&0.0198&0.0146&0.0086\\\hline
					KLL-0000 &0.0447&0.0299&0.0149&0.0063&0.0040\\\hline
					KLL-1000 &0.0321&0.0179&0.0095&0.0047&0.0025\\\hline
					KLL-1100 &0.0313&0.0166&0.0082&0.0043&0.0023\\\hline
					KLL-1110 &0.0322&0.0165&0.0084&0.0045&0.0023\\\hline
					KLL-1111 &0.0256&0.0146&0.0082&0.0043&0.0023\\\hline
		\end{tabular}}	
		\caption{}\label{fig:t1}
	\end{subfigure}	
	\begin{subfigure}[]{0.45\textwidth}
		{\small 
		\begin{tabular}{|l||*{5}{c|}}\hline
		\diagbox{ sketch}{ size}&$  2^7$&$2^8$&$2^9$&$2^{10}$&$2^{11}$ \\\hline\hline
	 				LWYC  	 &0.0117&0.0059&0.0026&0.0015&0.0008\\\hline
					KLL-0000 &0.0264&0.0104&0.0053&0.0017&0.0013\\\hline
					KLL-1000 &0.0107&0.0062&0.0028&0.0011&0.0008\\\hline
					KLL-1100 &0.0102&0.0067&0.0028&0.0014&0.0006\\\hline
					KLL-1110 &0.0121&0.0064&0.0034&0.0018&0.0008\\\hline
					KLL-1111 &0.0077&0.0043&0.0018&0.0008&0.0005\\\hline
		\end{tabular}}	
		\caption{}\label{fig:t2}
	\end{subfigure}	
	
	\caption{Figures \ref{fig:shuffle}, \ref{fig:sorted},  \ref{fig:caida}, and \ref{fig:wiki} depict the trade-off 
	between maximum error over all queried quantiles and the sketch size: Figures \ref{fig:shuffle} and  \ref{fig:sorted} 
	test the performance of the algorithms on shuffled and sorted data streams; Figures~\ref{fig:caida} and \ref{fig:wiki}
on CAIDA and Wikipedia datasets correspondingly. Tables \ref{fig:t1} and \ref{fig:t2} show the same trade-off, but make it possible 
to see the difference between different combos. Figure \ref{fig:constant} demonstrates independence of the algorithms performance 
from stream length, dashed lines indicate the sketch size equal 256 and the solid lines correspond to the sketch of size 1024. 
Finally, Figure~\ref{fig:trending} mix the trending data with a different ammounts of a random noise and demonstrates the influence of 
the stream order on the algorithm precision.}
%
\end{figure*}
\subsection{Implementation and Evaluation Details}
All the algorithms and experimental settings are implemented in Python 3.6.3.
The advantage of using a scripting language is fast prototyping and readable code for distribution inside the community.
Time performance of the algorithm is not the subject of the research in the current paper, and we leave its investigation for future work.
This in particular applies to the sweep compactor KLL and the algorithms for weighted quantiles, which theoretically improve the worst-case 
update time exponentially in $\epss$. 
All the algorithms in the current comparison are randomized, 
thus for each experiment the results presented are averaged over 50 independent runs. 
KLL and all suggested modifications are compared with each other and LWYC (the algorithm Random from \cite{huang2011sampling}). 
In \cite{wang2013quantiles} the authors carried on the experimental study of the algorithms from 
\cite{manku1998approximate,manku1999random, agarwal2013mergeable, greenwald2001space} and concluded that 
their own algorithm (LWYC) is preferable to the others: 
better in accuracy than \cite{greenwald2001space} and similar in accuracy compared with \cite{manku1999random} 
while LWYC has a simpler logic and easier to implement. 

As mentioned earlier we compared our algorithms under a fixed space restrictions. In other words, in all experiments we fixed the space 
allocated to the sketch and evaluated the algorithm based on the best accuracy it can achieve under that space limit. 
We measured the accuracy as the maximum deviation among all quantile queries, otherwise known as the Kolmogorov-Smirnov divergence, 
widely used to measure the distance between CDFs of two distributions. 
Additionally, we measure the introduced variance caused separately by the compaction steps and sampling.
Its value can help the user to evaluate the accuracy of the output.  
Note that for KLL this value depends on the size of the stream, and is independent of the arrival order of the items. 
In other words, the guarantees of KLL are the same for all types of streams, adversarial and structured. 
Some of our improvements change this property; recall that the sweep compactor KLL, when applied to sorted input, 
requires only a single sweep per layer. For this reason, in our experiments we found variance 
to be dependent not only on the internal randomness of the algorithm but also the arrival order of the stream items.

\subsection{Results} 

Note that the majority modifications presented in the current paper can be combined for better performance, 
due to the space limitations we present only some of them. For the sake of simplicity we will fix the order of 
suggested modification as: lazy from Section~\ref{sec:lazy}, reduced randomness from Section~\ref{sec:reduced_random}, 
error spreading from Section~\ref{sec:eq_spread} and sweeping from Section~\ref{sec:sweep}, 
and denote all possible combinations as four $0/1$ digits, i.e. $0000$ would imply the vanilla KLL without any
 modifications, while $0011$ would imply that we use KLL with error spreading trick and sweeping. 

 In Figures \ref{fig:shuffle} and \ref{fig:sorted} we compare the size/precision trade-off for LWYC, vanilla KLL, 
and KLL with  modifications. First, we can see that all KLL-based algorithms provide the approximation ratio significantly 
better than LWYC as the space allocation is growing, which confirms theoretical guarantees. Second, from the experiments it becomes
clear that all algorithms behave worse on the data without any order, i.e. shuffled stream. 
Although the laziness give the most significant push to the performance of the Vanilla KLL, all other modifications improve the precision even further if combined. One can 
 easily see it in the table \ref{fig:t1} for shuffled dataset and table \ref{fig:t2} for the sorted stream.
 Same experiments were carried on for the CAIDA dataset (Fig. \ref{fig:caida}), and shuffled Wikipedia page statistics (Fig. \ref{fig:wiki}).

Although, theoretically none of the algorithms should depend on the length of the dataset, 
we verified this property in practice, the results can be seen on Figure~\ref{fig:constant}.

In Figure \ref{fig:trending} we verified that although all the theoretical bounds hold, KLL and LWYC performance indeed 
depend on the ammount of randomness in the stream, more randomness leads to less precision. Our experiment were held on the 
trending dataset, i.e. the stream containing two components: $A\times$(mean-zero random variable) and $B\times$(trend $t/n$).
Figure  \ref{fig:trending} shows how precision drops as $A/B$ start to grow (X-axis). Note that modified algorithm does not 
drop in precision as fast as vanilla KLL or LWYC.

\section{High-Performance Implementation}\label{high-performance-section}

For simplicity of analysis, experimentation, and exposition, the
pseudocode so far assumes the use of list-based data structures. 
In reality those would include link fields that would double the space usage
for data types whose physical size is similar to that of pointers. 
Moreover, they are not very efficient in terms of update operations. 
In practice, a factor of two in space and update time is very significant. 

\begin{figure}
\vspace*{-0.5in}
\begin{center}
\hspace{3em}
\includegraphics[width=0.9\linewidth,angle=90]{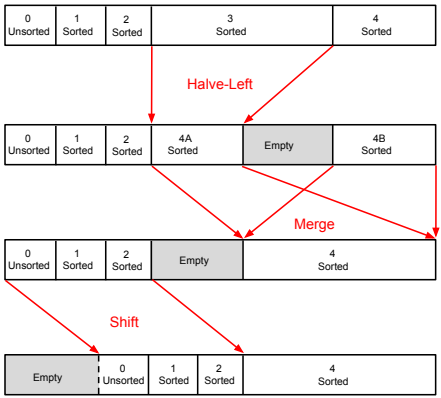}
\vspace*{-0.6in}
\caption{This diagram for Section \ref{high-performance-section} illustrates
the packed array data structure, and the data motion that occurs when a level is compacted.}
\label{data-motion-diagram}
\end{center}
\vspace{-0.2in}
\end{figure}

Fortunately, lazy KLL can be implemented in a way that optimizes the 
time and space constant factors. The key idea is to store the levels
in a shared and fully packed array, with the invariant that all levels
except for level 0 are already sorted. We note that
because there are no gaps between the subarrays occupied by the various
levels, the data motion that must occur during a compaction is somewhat tricky.
It is diagrammed in Figure~\ref{data-motion-diagram}. First, the algorithm
searches the levels from left to right to find the first one that is
at or above capacity. After a coin flip to decide between retaining the odd
or even positions, it then halves the items to the left and creates free space to the
right of the level. Then an in-place merge-reduce occurs with the level above (physically 
to the right). Finally, the data is shifted from all lower levels to close the gap created. 
We end up with free space on the extreme left of the array, that
can subsequently be used to grow the unsorted contents of level zero. 
Due to space restrictions we do not analyze this version of the algorithm here, but mention that the same asymptotic bound apply. 
An efficient implementation is available from the DataSketches \cite{datasketches} open source streaming algorithms library, the code can be found 
\url{datasketches.github.io} (in process of moving to \url{datasketches.apache.org})\footnote{The core code can be found at \url{github.com/apache/incubator-datasketches-java}}.

For completeness we provide a few run-time measurements of the high-performance implementation. Figure \ref{ocaml-timing-plot} (in the appendix) shows the results of stream-processing
timings that were performed on a 3.1 GHz MacBook Pro that had 16 GB of memory, and was running the Mojave operating system. The
initial behavior on each stream is somewhat complicated, but as the stream gets longer, the update time stabilizes at about 50 nanoseconds per item. We plotted the time without the sorting time for level 0 as well, to demonstrate that that part is significant in terms of speed.

\section{Conclusion}
We verified experimentally that the KLL algorithm proposed by Karnin~\etal~\cite{karnin2016optimal} has predicted asymptotic improvement over 
LWYC\cite{wang2013quantiles}.
We proposed four modifications to KLL with provably better constants in the approximation bounds. 
Experiments verified that the approximation is roughly twice as good in practice compared to KLL and more than four times better compared to LWYC (and growing with the space allocated to the sketch). 
Moreover, the worst-case update time for the presented sweep-compactor based KLL is $O(\log1/\eps)$ which improves over the rest of the compactor based algorithms.
Two algorithms proposed for the weighted streams improve over the naive extension from $O((\max{w_i}) \log1/\eps)$ to $O(\log 1/\eps)$
while maintaining the same space complexity. 
Finally, we provide an very efficient data structure for maintaining compactor based structures such as the algorithms above.


\bibliographystyle{ACM-Reference-Format}
\bibliography{bibliography}


\begin{thebibliography}{26}


\ifx \showCODEN    \undefined \def \showCODEN     #1{\unskip}     \fi
\ifx \showDOI      \undefined \def \showDOI       #1{#1}\fi
\ifx \showISBNx    \undefined \def \showISBNx     #1{\unskip}     \fi
\ifx \showISBNxiii \undefined \def \showISBNxiii  #1{\unskip}     \fi
\ifx \showISSN     \undefined \def \showISSN      #1{\unskip}     \fi
\ifx \showLCCN     \undefined \def \showLCCN      #1{\unskip}     \fi
\ifx \shownote     \undefined \def \shownote      #1{#1}          \fi
\ifx \showarticletitle \undefined \def \showarticletitle #1{#1}   \fi
\ifx \showURL      \undefined \def \showURL       {\relax}        \fi
\providecommand\bibfield[2]{#2}
\providecommand\bibinfo[2]{#2}
\providecommand\natexlab[1]{#1}
\providecommand\showeprint[2][]{arXiv:#2}

\bibitem[\protect\citeauthoryear{??}{cai}{2015}]%
        {caida}
 \bibinfo{year}{2015}\natexlab{}.
\newblock \showarticletitle{The CAIDA UCSD Anonymized Internet Traces,
  2015-02-19}.
\newblock  (\bibinfo{year}{2015}).
\newblock
\urldef\tempurl%
\url{http://www.caida.org/data/passive/passive_dataset.xml}
\showURL{%
\tempurl}


\bibitem[\protect\citeauthoryear{??}{wik}{2016}]%
        {wiki}
 \bibinfo{year}{2016}\natexlab{}.
\newblock \showarticletitle{Page view statistics for Wikimedia projects}.
\newblock  (\bibinfo{year}{2016}).
\newblock
\urldef\tempurl%
\url{https://dumps.wikimedia.org/other/pagecounts-raw/}
\showURL{%
\tempurl}


\bibitem[\protect\citeauthoryear{Agarwal, Cormode, Huang, Phillips, Wei, and
  Yi}{Agarwal et~al\mbox{.}}{2013}]%
        {agarwal2013mergeable}
\bibfield{author}{\bibinfo{person}{Pankaj~K Agarwal}, \bibinfo{person}{Graham
  Cormode}, \bibinfo{person}{Zengfeng Huang}, \bibinfo{person}{Jeff~M
  Phillips}, \bibinfo{person}{Zhewei Wei}, {and} \bibinfo{person}{Ke Yi}.}
  \bibinfo{year}{2013}\natexlab{}.
\newblock \showarticletitle{Mergeable summaries}.
\newblock \bibinfo{journal}{\emph{ACM Transactions on Database Systems (TODS)}}
  \bibinfo{volume}{38}, \bibinfo{number}{4} (\bibinfo{year}{2013}),
  \bibinfo{pages}{26}.
\newblock


\bibitem[\protect\citeauthoryear{Arasu and Manku}{Arasu and Manku}{2004}]%
        {arasu2004approximate}
\bibfield{author}{\bibinfo{person}{Arvind Arasu} {and}
  \bibinfo{person}{Gurmeet~Singh Manku}.} \bibinfo{year}{2004}\natexlab{}.
\newblock \showarticletitle{Approximate counts and quantiles over sliding
  windows}. In \bibinfo{booktitle}{\emph{Proceedings of the twenty-third ACM
  SIGMOD-SIGACT-SIGART symposium on Principles of database systems}}. ACM,
  \bibinfo{pages}{286--296}.
\newblock


\bibitem[\protect\citeauthoryear{Chen and Guestrin}{Chen and Guestrin}{2016}]%
        {chen2016xgboost}
\bibfield{author}{\bibinfo{person}{Tianqi Chen} {and} \bibinfo{person}{Carlos
  Guestrin}.} \bibinfo{year}{2016}\natexlab{}.
\newblock \showarticletitle{Xgboost: A scalable tree boosting system}. In
  \bibinfo{booktitle}{\emph{Proceedings of the 22nd acm sigkdd international
  conference on knowledge discovery and data mining}}. ACM,
  \bibinfo{pages}{785--794}.
\newblock


\bibitem[\protect\citeauthoryear{Cormode, Garofalakis, Muthukrishnan, and
  Rastogi}{Cormode et~al\mbox{.}}{2005}]%
        {cormode2005holistic}
\bibfield{author}{\bibinfo{person}{Graham Cormode}, \bibinfo{person}{Minos
  Garofalakis}, \bibinfo{person}{S Muthukrishnan}, {and}
  \bibinfo{person}{Rajeev Rastogi}.} \bibinfo{year}{2005}\natexlab{}.
\newblock \showarticletitle{Holistic aggregates in a networked world:
  Distributed tracking of approximate quantiles}. In
  \bibinfo{booktitle}{\emph{Proceedings of the 2005 ACM SIGMOD international
  conference on Management of data}}. ACM, \bibinfo{pages}{25--36}.
\newblock


\bibitem[\protect\citeauthoryear{DeWitt, Naughton, and Schneider}{DeWitt
  et~al\mbox{.}}{1991}]%
        {dewitt1991parallel}
\bibfield{author}{\bibinfo{person}{David~J DeWitt}, \bibinfo{person}{Jeffrey~F
  Naughton}, {and} \bibinfo{person}{Donovan~A Schneider}.}
  \bibinfo{year}{1991}\natexlab{}.
\newblock \showarticletitle{Parallel sorting on a shared-nothing architecture
  using probabilistic splitting}. In \bibinfo{booktitle}{\emph{Parallel and
  distributed information systems, 1991., proceedings of the first
  international conference on}}. IEEE, \bibinfo{pages}{280--291}.
\newblock


\bibitem[\protect\citeauthoryear{Felber and Ostrovsky}{Felber and
  Ostrovsky}{2015}]%
        {felber2015randomized}
\bibfield{author}{\bibinfo{person}{David Felber} {and} \bibinfo{person}{Rafail
  Ostrovsky}.} \bibinfo{year}{2015}\natexlab{}.
\newblock \showarticletitle{A randomized online quantile summary in O
  (1/epsilon* log (1/epsilon)) words}. In
  \bibinfo{booktitle}{\emph{LIPIcs-Leibniz International Proceedings in
  Informatics}}, Vol.~\bibinfo{volume}{40}. Schloss Dagstuhl-Leibniz-Zentrum
  fuer Informatik.
\newblock


\bibitem[\protect\citeauthoryear{Greenwald and Khanna}{Greenwald and
  Khanna}{2001}]%
        {greenwald2001space}
\bibfield{author}{\bibinfo{person}{Michael Greenwald} {and}
  \bibinfo{person}{Sanjeev Khanna}.} \bibinfo{year}{2001}\natexlab{}.
\newblock \showarticletitle{Space-efficient online computation of quantile
  summaries}. In \bibinfo{booktitle}{\emph{ACM SIGMOD Record}},
  Vol.~\bibinfo{volume}{30}. ACM, \bibinfo{pages}{58--66}.
\newblock


\bibitem[\protect\citeauthoryear{Greenwald and Khanna}{Greenwald and
  Khanna}{2004}]%
        {greenwald2004power}
\bibfield{author}{\bibinfo{person}{Michael~B Greenwald} {and}
  \bibinfo{person}{Sanjeev Khanna}.} \bibinfo{year}{2004}\natexlab{}.
\newblock \showarticletitle{Power-conserving computation of order-statistics
  over sensor networks}. In \bibinfo{booktitle}{\emph{Proceedings of the
  twenty-third ACM SIGMOD-SIGACT-SIGART symposium on Principles of database
  systems}}. ACM, \bibinfo{pages}{275--285}.
\newblock


\bibitem[\protect\citeauthoryear{Greenwald and Khanna}{Greenwald and
  Khanna}{2016}]%
        {greenwald2016quantiles}
\bibfield{author}{\bibinfo{person}{Michael~B Greenwald} {and}
  \bibinfo{person}{Sanjeev Khanna}.} \bibinfo{year}{2016}\natexlab{}.
\newblock \showarticletitle{Quantiles and equi-depth histograms over streams}.
\newblock In \bibinfo{booktitle}{\emph{Data Stream Management}}.
  \bibinfo{publisher}{Springer}, \bibinfo{pages}{45--86}.
\newblock


\bibitem[\protect\citeauthoryear{Huang, Wang, Yi, and Liu}{Huang
  et~al\mbox{.}}{2011}]%
        {huang2011sampling}
\bibfield{author}{\bibinfo{person}{Zengfeng Huang}, \bibinfo{person}{Lu Wang},
  \bibinfo{person}{Ke Yi}, {and} \bibinfo{person}{Yunhao Liu}.}
  \bibinfo{year}{2011}\natexlab{}.
\newblock \showarticletitle{Sampling based algorithms for quantile computation
  in sensor networks}. In \bibinfo{booktitle}{\emph{Proceedings of the 2011 ACM
  SIGMOD International Conference on Management of data}}. ACM,
  \bibinfo{pages}{745--756}.
\newblock


\bibitem[\protect\citeauthoryear{Karnin, Lang, and Liberty}{Karnin
  et~al\mbox{.}}{2016}]%
        {karnin2016optimal}
\bibfield{author}{\bibinfo{person}{Zohar Karnin}, \bibinfo{person}{Kevin Lang},
  {and} \bibinfo{person}{Edo Liberty}.} \bibinfo{year}{2016}\natexlab{}.
\newblock \showarticletitle{Optimal quantile approximation in streams}. In
  \bibinfo{booktitle}{\emph{Foundations of Computer Science (FOCS), 2016 IEEE
  57th Annual Symposium on}}. IEEE, \bibinfo{pages}{71--78}.
\newblock


\bibitem[\protect\citeauthoryear{Li, Li, Wang, and Cao}{Li
  et~al\mbox{.}}{2011}]%
        {li2011ubiquitous}
\bibfield{author}{\bibinfo{person}{Zhenjiang Li}, \bibinfo{person}{Mo Li},
  \bibinfo{person}{Jiliang Wang}, {and} \bibinfo{person}{Zhichao Cao}.}
  \bibinfo{year}{2011}\natexlab{}.
\newblock \showarticletitle{Ubiquitous data collection for mobile users in
  wireless sensor networks}. In \bibinfo{booktitle}{\emph{INFOCOM, 2011
  Proceedings IEEE}}. IEEE, \bibinfo{pages}{2246--2254}.
\newblock


\bibitem[\protect\citeauthoryear{Lin, Lu, Xu, and Yu}{Lin
  et~al\mbox{.}}{2004}]%
        {lin2004continuously}
\bibfield{author}{\bibinfo{person}{Xuemin Lin}, \bibinfo{person}{Hongjun Lu},
  \bibinfo{person}{Jian Xu}, {and} \bibinfo{person}{Jeffrey~Xu Yu}.}
  \bibinfo{year}{2004}\natexlab{}.
\newblock \showarticletitle{Continuously maintaining quantile summaries of the
  most recent n elements over a data stream}. In \bibinfo{booktitle}{\emph{Data
  Engineering, 2004. Proceedings. 20th International Conference on}}. IEEE,
  \bibinfo{pages}{362--373}.
\newblock


\bibitem[\protect\citeauthoryear{Liu, Manousis, Vorsanger, Sekar, and
  Braverman}{Liu et~al\mbox{.}}{2016}]%
        {liu2016one}
\bibfield{author}{\bibinfo{person}{Zaoxing Liu}, \bibinfo{person}{Antonis
  Manousis}, \bibinfo{person}{Gregory Vorsanger}, \bibinfo{person}{Vyas Sekar},
  {and} \bibinfo{person}{Vladimir Braverman}.} \bibinfo{year}{2016}\natexlab{}.
\newblock \showarticletitle{One sketch to rule them all: Rethinking network
  flow monitoring with univmon}. In \bibinfo{booktitle}{\emph{Proceedings of
  the 2016 ACM SIGCOMM Conference}}. ACM, \bibinfo{pages}{101--114}.
\newblock


\bibitem[\protect\citeauthoryear{Manku, Rajagopalan, and Lindsay}{Manku
  et~al\mbox{.}}{1998}]%
        {manku1998approximate}
\bibfield{author}{\bibinfo{person}{Gurmeet~Singh Manku},
  \bibinfo{person}{Sridhar Rajagopalan}, {and} \bibinfo{person}{Bruce~G
  Lindsay}.} \bibinfo{year}{1998}\natexlab{}.
\newblock \showarticletitle{Approximate medians and other quantiles in one pass
  and with limited memory}. In \bibinfo{booktitle}{\emph{ACM SIGMOD Record}},
  Vol.~\bibinfo{volume}{27}. ACM, \bibinfo{pages}{426--435}.
\newblock


\bibitem[\protect\citeauthoryear{Manku, Rajagopalan, and Lindsay}{Manku
  et~al\mbox{.}}{1999}]%
        {manku1999random}
\bibfield{author}{\bibinfo{person}{Gurmeet~Singh Manku},
  \bibinfo{person}{Sridhar Rajagopalan}, {and} \bibinfo{person}{Bruce~G
  Lindsay}.} \bibinfo{year}{1999}\natexlab{}.
\newblock \showarticletitle{Random sampling techniques for space efficient
  online computation of order statistics of large datasets}. In
  \bibinfo{booktitle}{\emph{ACM SIGMOD Record}}, Vol.~\bibinfo{volume}{28}.
  ACM, \bibinfo{pages}{251--262}.
\newblock


\bibitem[\protect\citeauthoryear{Munro and Paterson}{Munro and
  Paterson}{1980}]%
        {munro1980selection}
\bibfield{author}{\bibinfo{person}{J~Ian Munro} {and} \bibinfo{person}{Mike~S
  Paterson}.} \bibinfo{year}{1980}\natexlab{}.
\newblock \showarticletitle{Selection and sorting with limited storage}.
\newblock \bibinfo{journal}{\emph{Theoretical computer science}}
  \bibinfo{volume}{12}, \bibinfo{number}{3} (\bibinfo{year}{1980}),
  \bibinfo{pages}{315--323}.
\newblock


\bibitem[\protect\citeauthoryear{Pike, Dorward, Griesemer, and Quinlan}{Pike
  et~al\mbox{.}}{2005}]%
        {pike2005interpreting}
\bibfield{author}{\bibinfo{person}{Rob Pike}, \bibinfo{person}{Sean Dorward},
  \bibinfo{person}{Robert Griesemer}, {and} \bibinfo{person}{Sean Quinlan}.}
  \bibinfo{year}{2005}\natexlab{}.
\newblock \showarticletitle{Interpreting the data: Parallel analysis with
  Sawzall}.
\newblock \bibinfo{journal}{\emph{Scientific Programming}}
  \bibinfo{volume}{13}, \bibinfo{number}{4} (\bibinfo{year}{2005}),
  \bibinfo{pages}{277--298}.
\newblock


\bibitem[\protect\citeauthoryear{Poosala, Haas, Ioannidis, and Shekita}{Poosala
  et~al\mbox{.}}{1996}]%
        {poosala1996improved}
\bibfield{author}{\bibinfo{person}{Viswanath Poosala}, \bibinfo{person}{Peter~J
  Haas}, \bibinfo{person}{Yannis~E Ioannidis}, {and} \bibinfo{person}{Eugene~J
  Shekita}.} \bibinfo{year}{1996}\natexlab{}.
\newblock \showarticletitle{Improved histograms for selectivity estimation of
  range predicates}. In \bibinfo{booktitle}{\emph{ACM Sigmod Record}},
  Vol.~\bibinfo{volume}{25}. ACM, \bibinfo{pages}{294--305}.
\newblock


\bibitem[\protect\citeauthoryear{Rhodes, Lang, Saydakov, Liberty, and
  Thaler}{Rhodes et~al\mbox{.}}{2013}]%
        {datasketches}
\bibfield{author}{\bibinfo{person}{Lee Rhodes}, \bibinfo{person}{Kevin Lang},
  \bibinfo{person}{Alexander Saydakov}, \bibinfo{person}{Edo Liberty}, {and}
  \bibinfo{person}{Justin Thaler}.} \bibinfo{year}{2013}\natexlab{}.
\newblock \bibinfo{title}{DataSketches: A library of stochastic streaming
  algorithms}.
\newblock \bibinfo{howpublished}{Open source software:
  https://datasketches.github.io/, (in process of moving to
  \url{datasketches.apache.org})}.   (\bibinfo{year}{2013}).
\newblock


\bibitem[\protect\citeauthoryear{Selinger, Astrahan, Chamberlin, Lorie, and
  Price}{Selinger et~al\mbox{.}}{1979}]%
        {selinger1979access}
\bibfield{author}{\bibinfo{person}{P~Griffiths Selinger},
  \bibinfo{person}{Morton~M Astrahan}, \bibinfo{person}{Donald~D Chamberlin},
  \bibinfo{person}{Raymond~A Lorie}, {and} \bibinfo{person}{Thomas~G Price}.}
  \bibinfo{year}{1979}\natexlab{}.
\newblock \showarticletitle{Access path selection in a relational database
  management system}. In \bibinfo{booktitle}{\emph{Proceedings of the 1979 ACM
  SIGMOD international conference on Management of data}}. ACM,
  \bibinfo{pages}{23--34}.
\newblock


\bibitem[\protect\citeauthoryear{Shrivastava, Buragohain, Agrawal, and
  Suri}{Shrivastava et~al\mbox{.}}{2004}]%
        {shrivastava2004medians}
\bibfield{author}{\bibinfo{person}{Nisheeth Shrivastava},
  \bibinfo{person}{Chiranjeeb Buragohain}, \bibinfo{person}{Divyakant Agrawal},
  {and} \bibinfo{person}{Subhash Suri}.} \bibinfo{year}{2004}\natexlab{}.
\newblock \showarticletitle{Medians and beyond: new aggregation techniques for
  sensor networks}. In \bibinfo{booktitle}{\emph{Proceedings of the 2nd
  international conference on Embedded networked sensor systems}}. ACM,
  \bibinfo{pages}{239--249}.
\newblock


\bibitem[\protect\citeauthoryear{Wang, Luo, Yi, and Cormode}{Wang
  et~al\mbox{.}}{2013}]%
        {wang2013quantiles}
\bibfield{author}{\bibinfo{person}{Lu Wang}, \bibinfo{person}{Ge Luo},
  \bibinfo{person}{Ke Yi}, {and} \bibinfo{person}{Graham Cormode}.}
  \bibinfo{year}{2013}\natexlab{}.
\newblock \showarticletitle{Quantiles over data streams: an experimental
  study}. In \bibinfo{booktitle}{\emph{Proceedings of the 2013 ACM SIGMOD
  International Conference on Management of Data}}. ACM,
  \bibinfo{pages}{737--748}.
\newblock


\bibitem[\protect\citeauthoryear{Yi and Zhang}{Yi and Zhang}{2013}]%
        {yi2013optimal}
\bibfield{author}{\bibinfo{person}{Ke Yi} {and} \bibinfo{person}{Qin Zhang}.}
  \bibinfo{year}{2013}\natexlab{}.
\newblock \showarticletitle{Optimal tracking of distributed heavy hitters and
  quantiles}.
\newblock \bibinfo{journal}{\emph{Algorithmica}} \bibinfo{volume}{65},
  \bibinfo{number}{1} (\bibinfo{year}{2013}), \bibinfo{pages}{206--223}.
\newblock


\end{thebibliography}

\newpage
~
\newpage
~
\appendix


\section{Fixing the Original KLL Proof} \label{app:kllfix}
The original paper by Karnin~\etal~\cite{karnin2016optimal} contains a mistake regarding the number of compactions performed at a single level. 
Correcting the mistake is trivial and does not change the authors claim. 
Nevertheless, we provide a correction of their argument.
The authors use compactors of exponentially decreasing size. 
Higher weight items receive higher capacity compactors. 
The error appeared in the last inequality of the bound on~$m_h$ --- the number of compaction made at level $h$ 
(page $6$ in \cite{karnin2016optimal}):
\begin{equation} \label{eq:kllProofIssue}
m_h \le \frac{n}{k_hw_h} \le \frac{2n}{k2^H} (2/c)^{H-h} \le (2/c)^{H-h-1},
\end {equation}
where $H$ is the height of the top compactor, $k_h = kc^{H-h}$ is the size of the compactor at height $h$. 
Note, that the last inequality implies $n \le ck2^{H-2} = k_{h-1}w_{h-1}$, while from the defition of $H$ it follows 
that at least one compaction happened on level $H-1$. Therefore $n \ge k_{h-1}w_{h-1}$. 
Fixing this slightly increases the constant in the final upper bound. 

Recall that $k\ge4$ and $c\in(0.5,1)$. 
We reuse the notation and refer to the height of the top compactor as $H$.
Additionally, we introduce $H'$ which denotes the height of the top compactor of size $2$. 
Due to the choice of $k$ and $c$ we can conclude that $H' \le H -1$. 

Every compactor of size $2$ contains at most one item, otherwise it would be compacted. 
Therefore, the bottom $H'$ compactors have total weight $\sum_{h=1}^{H'}{w_h} = \sum_{h=1}^{H'}{2^{h-1}} \le 2^{H'}$.
Similarly, every compactor of size $k_h$ contains at most $k_h-1 = kc^{H-h} -1 \le (k-1)c^{H-h}$ items.
Then the total weight of compactors from level $H'+1$ to $H$ is: 
$$\sum_{h=H'+1}^{H}{\hspace{-7pt}(k_h-1)w_h} \le \sum_{h=1}^{H}{(k-1)c^{H-h}2^{h-1}}=(k-1)c^{H-1}\sum_{h=1}^{H}{\left(2/c\right)^{h-1}}$$
$$= (k-1)c^{H-1}\frac{(2/c)^H - 1}{2/c-1}\le \frac{(k-1)2^H}{2-c}\le (k-1)2^H $$
Putting together the total weight of the bottom $H'$ and top $H-H'$ compactors we get the upper bound on the number of items processed:
$$n \le (k-1)2^H + 2^{H'} \le (k - 1 + 1/2)2^H \le k2^H.$$
Plugging $n \le k2^H$ into the last inequality of Equation~\ref{eq:kllProofIssue} leads to $m_h \le 2 (2/c)^{H-h}$
which is $4/c$ times worse than the initial derivation.
Repeating the argument as in \cite{karnin2016optimal} and in the Section \ref{section:unifiedview} of the current paper, 
we get $\sum_{h=1}^H\sum_{i=1}^{m_h}w_h^2 \le \frac{2n^2/k^2}{c^3(2c-1)}$.
As in \cite{karnin2016optimal} applying Hoeffding's inequality gives
$$P(|\text{Err}|> \eps n) \le 2\exp{\left(-C\eps^2k^2\right)} \le \delta$$. 
However, the constant $C$ has changed from $2c^2(2c-1)$ to $C = \frac{1}{2}c^3(2c-1)$. 
Note that all asymptotic guarantees stay the same as in \cite{karnin2016optimal}.


\section{Missing Proofs} \label{app:missing proofs}

\noindent {\bf Theorem~\ref{thm:base2update}}
Algorithm~\ref{code:base2update} processes a stream of weighted updates.
It outputs all $\eps$-approximate quantiles with high probability using memory $O(\epss\sqrt{\log1/\eps})$. 
In the worst-case scenario, a single update invokes $O(\log1/\eps)$ update calls to compactors of the KLL sketch 
and $O(\epss\sqrt{\log1/\eps})$ calls to the sampler. 
This results in a $O(\epss\log^{3/2}\epss)$ worst-case update run-time. 
The amortized run-time is $O(\log^2\epss)$.

\begin{proof}
The analysis of the error of this algorithm is straightforward, as an item of weight $w$ is broken into several weights summing to $w$. 
For the runtime analysis, we decompose it into three parts:
\begin{enumerate}
	\item increase of $H$ (lines \ref{codelines:base2update:1}-\ref{codelines:base2update:2} in Algorithm \ref{code:base2update})
	\item push $w'$ to sampler and all $a_h$ to levels $h<H$ (lines \ref{codelines:base2update:3},\ref{codelines:base2update:4})
	\item push $a_H$ to the top compactor (line \ref{codelines:base2update:5})
\end{enumerate}
For the first part, the worst-case happens when all compactors are full and all should be deleted due to increase of $H$.
Therefore, line~\ref{codelines:base2update:12} of Algorithm~\ref{code:base2update} will push $O(\epss\sqrt{\log\epss})$ items into the data 
structure at the total cost $O(\epss\log^{3/2} \epss)$.
Since each of these items must have been inserted earlier, the amortized runtime for the first part is $O(\log \epss)$.

The second part is associated with the different components of $w$ except the largest one, $a_H$.
In the worst-case, $a_h=1$ for all $ h \in (H_s, H)$. 
Recall, that the worst-case runtime of lazy sweeping KLL is $O(\log\epss)$.
Therefore the worst-case runtime is $O((H - H_s)\log\epss)= O(\log^2\epss)$. 
The amortized running time is $O(\log^2\epss)$ as well.

Finally, the third part is taking into account adding the same element to the top layer $a_H$ times.
In the worst-case, in total $O(k\log\epss) = O(\epss\log^{3/2}\epss)$ time. 
For the amortized case, although $a_H$ could be equal to $k-1$, we do not really need to add $a_H$ copies 
of the item but rather remember the number of times the item is inserted. 
It follows that the amortized case is the same as that of inserting an item to a compactor which is $O(\log(k))=O(\log(\epss))$

Summing the three components, we get a worst-case runtime of $O(\epss\log^{3/2}\epss)$ and amortized time of $O(\log^2\epss)$.
\end{proof}


\noindent {\bf Lemma~\ref{lemma:base2compactor} }
Given a stream of $n$ items of weights in $[w,2w)$, a weight-aware compactor outputs a stream of $n/2$ items of weight $[2w,4w)$. 
For a size $k$ weight-aware compactor the added error in rank between input and output stream (for any query $a$) is equal to $\sum_{i=1}^{n/k} X_i$. 
Here, $X_i$'s are independent random variable such that $\text{E}[X_i]=0$ and $|X_i| < 2w$.

\begin{proof}
The claim regarding the stream length is trivial as every two items become a single item in the compact operation. Also, since the weight of an output is $w_a + w_b$, with $w_a,w_b \in [w,2w)$ it follows that the output weights are in $[2w,4w)$. For the error, consider an arbitrary query $q$. In a single compact operation, $q$ is an inner query if $a_j < q < a_{j+1}$ for some even $j$ and an outer query otherwise. If $q$ is an outer query, the error associated to it is 0. Otherwise, the error is $w_b$ with probability $\frac{w_a}{w_a + w_b}$ and $-w_a$ with probability $\frac{w_b}{w_a + w_b}$. Denoting the error for $q$ at compaction $i$ as $X_i$ we get that $\text{E}[X_i]=0$ and $|X_i| < 2w$ as claimed. 
Finally, since the size of the compactor is $k$, a compact operation will occur for every $k$ items and indeed the number of error variables $X_i$ is $n/k$.
\end{proof}

\noindent {\bf Theorem~\ref{thm:base2compactor} }
Algorithm \ref{code:base2compactor} processes a stream of weighted updates and outputs all $\eps$-approximate 
quantiles with high probability using space $O(\epss\sqrt{\log1/\eps})$. It's worst-case update time is $O(\log1/\eps)$. 

\begin{proof}
The error in the algorithm is introduced via compaction procedures in line~\ref{codelines:base2compactor:2} and via dropping bottom compactors in 
line~\ref{codelines:base2compactor:1}. 
The analysis for this compactor extension is similar to the unitary weighted case.
Let $X_{i,h}$ be a random variable which indicates the sign of the error introduced 
during the $i$-th compaction on the $h$-th level, and let it be equal to zero if no error is introduced. 
Note, that in Lemma~\ref{lemma:base2compactor} we showed that $E(X_{i,h}) = 0$ and $X_{i,h}\le 2w$, therefore the total error introduced is 
$$\text{Err} = \sum_{h=1}^H{\sum_{i=1}^{m_h}{2w_hX_{i,h}}}. $$
Repeating the argument as in Appendix~\ref{app:kllfix} we conclude that:
$$P(|\text{Err}| > \eps W)\le 2\exp{(-\frac{C}{4}\eps^2k^2)}< \delta.$$
To reach the same approximation guarantees with the same probability of failure, 
one need to set $k_{new} = 2k_{old}$, i.e.\ this algorithm will use twice as much space as the naive implementations. 
Additionally it stores a weight for each item explicitly which doubles the space complexity (this depends on the memory footprint of a stream item). 

Note, that the error introduced in line~\ref{codelines:base2compactor:1} is not $0$ in expectation and might accumulate over time.
However, line~\ref{codelines:base2compactor:1} is only executed when an item of weight more than $k2^H$ is processed. 
We can bound the overall weight of the items in the bottom compactors that were discarded as a small fraction of the overall number of items processed. The cumulative weight of items dropped will turn out to be a geometric sequence dominated by its last element, which in turn is a small fraction of the overall weight. 
In Appendix~\ref{app:kllfix} we show that the total  weight of items in compactor of height $h$ is at most $k2^h$
For weighted compactors it is $k2^{h+1}$.
The number of bottom compactors that are to be dropped is $H_{\text{new}}- H - 1$ and the level of the highest dropped compactor  is 
$H_s + H_{\text{new}} - H -1$. 
Therefore, the total weight dropped is less than $k2^{H_s + H_{\text{new}} - H -1}$.
Our goal is to bound the portion of total weight we dropped. 
Therefore, we will estimate the ratio of dropped weight to the weight of added items.
$$ \frac{k2^{H_s + H_{\text{new}} - H -1}}{k2^{ H_{\text{new}} - 1}}=2^{ H_s - H} = 2^{-\log_{1/c} k} = k^{- 1/\log_2{1/c}}\le \eps$$
The last equation holds since $c>0.5$ and $\eps=\Omega(1/k\sqrt{\log(k)})$.
It follows that with each compactor drop we discard at most an $\eps$ portion of the stream. 
At the same time before every such drop  
the total weight increase by at least $50\%$: $\frac{\text {added weight}}{\text{current weight}} \ge \frac{k2^{H}}{k2^{H+1}}\ge 0.5$.
We conclude that if we have a large number of compactor drops, the final error introduced is 
$\eps(1 +(2/3) + (2/3)^2 + \ldots) \le 3\eps$. 
Adjusting the input memory allowance by a constant factor leads to the desired approximation.

To process any weighted update, Algorithm~\ref{code:base2compactor} applies lines \ref{codelines:base2compactor:2}, 
\ref{codelines:base2compactor:3} and \ref{codelines:base2compactor:4}. If we use lazy compactions with sweeping, 
lines \ref{codelines:base2compactor:2} and \ref{codelines:base2compactor:3} in the worst case require  $O(\log\epss)$ running time.
As for line \ref{codelines:base2compactor:4}, we store a single item $(a, 2^H)$ and its multiplicity $w/2^H \le k$, instead of 
pushing up to $k$ items into the top compactor. Hence, in the worst case line \ref{codelines:base2compactor:4} accounts for $O(1)$ run-time.  
\end{proof}

\section{Base2update update procedure}
Algorithm~\ref{code:base2compactor} contains the pseudo-code for the Base2update update procedure. In it, \emph{rb} is a uniform random number in $[0,1]$, and append,delete,sort are the standard operations of a list.
\begin{algorithm}
	\caption{Base2update update procedure}\label{code:base2compactor}
\begin{algorithmic}[1]
\small
	\Function{update}{$(a,w)$}
		\State  $H_{\text{new}} = \argmin_h(2^h > w/k)$
		\If{$H_{\text{new}} > H$}
		\State delete bottom $H_{\text{new}} - H$ compactors \label{codelines:base2compactor:1}
			\State add $H_{\text{new}} - H$ empty compactors on the top; $H = H_{\text{new}}$
		\EndIf 
		\State  $h = \argmin_h(2^h > w)$	
		\State\algorithmicif\ {$h \le H_{s}$}
		\State \hspace{\algorithmicindent} \algorithmicif\ \Call{Sampler}{$(a,w)$} \algorithmicthen\ \Call{KLL$[0]$.append}{$(a, 2^{H_s})$}; itemsN++  \label{codelines:base2compactor:2}		
 		\State \algorithmicelse\algorithmicif\ {$h \le H$}
		\State \hspace{\algorithmicindent} \Call{KLL$[h]$.append}{$(a, w)$}; itemsN++     \label{codelines:base2compactor:3}
		\State \algorithmicelse
		\State \hspace{\algorithmicindent} $w/2^H < k$ times \textbf{repeat:} \Call{KLL$[H]$.append}{$(a, 2^H)$}; itemsN++     
		\label{codelines:base2compactor:4}
		\If{itemsN $>$ sketchSpace}
	    	\For{$h = 1\ldots H$}
		\State \algorithmicif { \Call{len}{KLL[$h$]} $\ge$ $k_h$}\; \Call{KLL.compact}{$h$};\; \textbf{break}
	    	\EndFor
		\EndIf
	\EndFunction
	\Function{KLL.compact}{h} \Comment{lazy sweeping}
		\State \Call{KLL[$h$].sort}{}(); r $= $\Call{random}{$[0,1]$};
		\State find smallest $j$, s.t. $a_j \ge \theta;\;\;\; w = w_j + w_{j+1}$ 
		\State\algorithmicif\ {rb $\le w_j/w$} \algorithmicthen\ $a = a_j$ \algorithmicelse\ $a = a_{j+1}$
		\State\Call{KLL[$h+1$].append}{(a,w)}; \Call{KLL[$h$].delete}{$a_j, a_{j+1}$}; 
	\EndFunction
\end{algorithmic}
\end{algorithm}

\section{Speed measurements of efficient implementation}

\begin{figure}[ht]
\begin{center}
\includegraphics[width=0.7\linewidth]{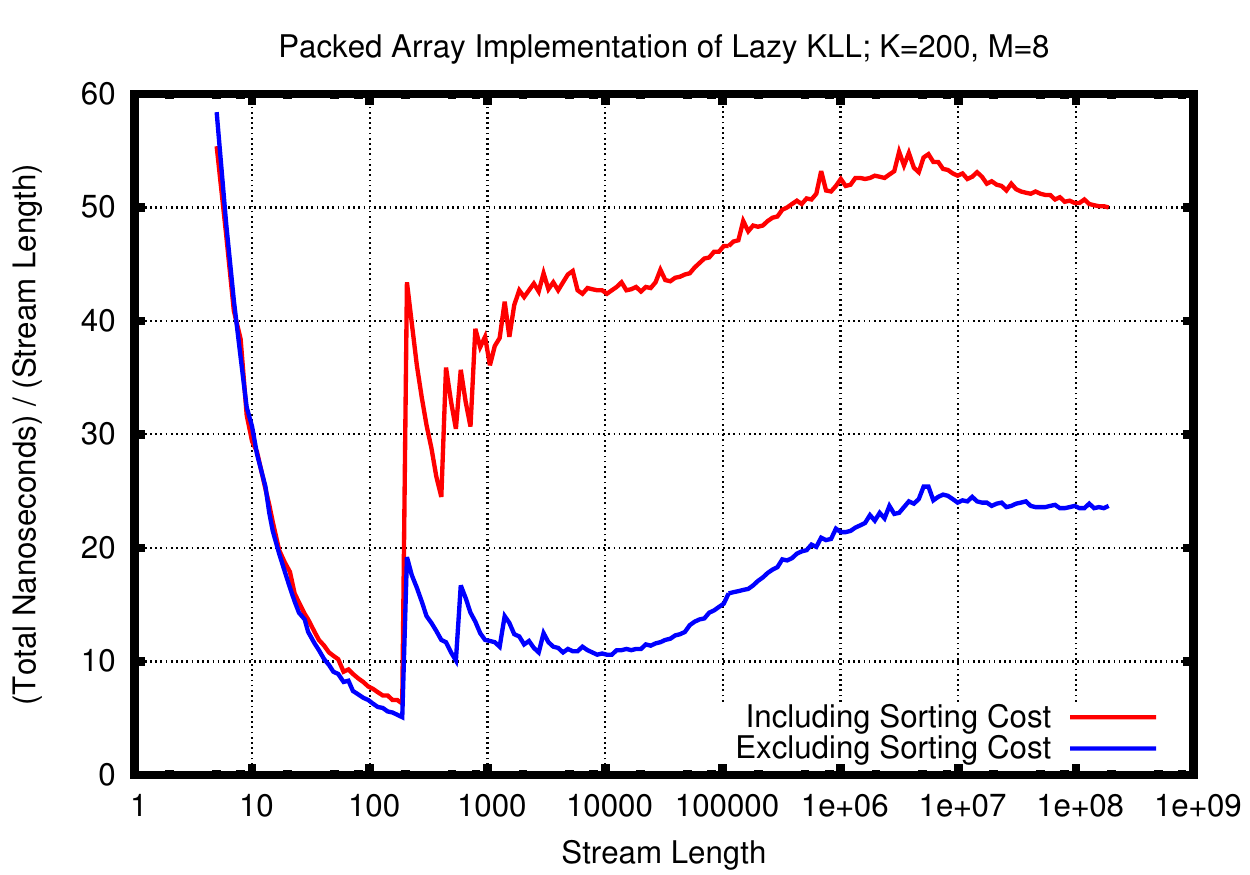}
\caption{Average update time per item in nanoseconds for the lazy KLL algorithm described in Section \ref{high-performance-section}.
After the sketch has filled up, the algorithm spends roughly half its time sorting items in the level 0 buffer.}
\label{ocaml-timing-plot}
\end{center}
\end{figure}

Figure~\ref{ocaml-timing-plot} plots the runtime measurements of the high performance implementation discussed in Section~\ref{high-performance-section}.
For each of
several values of $n$ we measured the average (over multiple trials)
time required to process a stream that consisted of a random
permutation of the integers between 1 and $n$. The $y$-axis of the
plot is the total processing time (including object creation and
initialization, and garbage collection), measured in nanoseconds, and
then divided by $n$. This can be interpreted as the average per-item update
time for the data structure. The blue line excludes the sorting of level 0 and the red line includes the entire procedure.
Experiments were conducted using a single thread and an Intel Core I7-2670qm 3.1 Ghz processor.

\end{document}